\documentclass[epj]{svjour}

\usepackage[utf8]{inputenc}

\usepackage{amsmath}
\usepackage{amssymb}
\usepackage{wasysym}

\usepackage{siunitx}
\DeclareSIUnit\mearth{M_\oplus}
\DeclareSIUnit\rearth{R_\oplus}
\DeclareSIUnit\mj{M_{\textrm{\jupiter}}}
\DeclareSIUnit\rj{R_{\textrm{\jupiter}}}
\DeclareSIUnit\msun{M_\odot}
\DeclareSIUnit\year{yr}
\DeclareSIUnit\au{au}

\usepackage{natbib}

\usepackage{graphicx}
\usepackage{url}
\usepackage{hyperref}
\hypersetup{
    colorlinks = true,
    linkcolor = {blue},
    citecolor = {blue},
    urlcolor = {blue}
}

\begin{document}

\title{Planetary Population Synthesis and the Emergence of Four Classes of Planetary System Architectures}
\author{Alexandre Emsenhuber\inst{1,2,3,}\thanks{emsenhuber@usm.lmu.de} \and Christoph Mordasini\inst{2} \and Remo Burn\inst{4}}

\institute{Universitäts-Sternwarte, Ludwig-Maximilians-Universität München, Scheinerstraße 1, 81679 München, Germany \and Space Research and Planetary Sciences, Universität Bern, Gesellschaftsstrasse 6, 3012 Bern, Switzerland \and Lunar and Planetary Laboratory, University of Arizona, 1629 E. University Blvd., Tucson, AZ 85721, USA \and Max-Planck-Institut für Astronomie, Königstuhl 17, 69117, Heidelberg, Germany}

\abstract{Planetary population synthesis is a helpful tool to understand the physics of planetary system formation. It builds on a global model, meaning that the model has to include a multitude of physical processes. The outcome can be statistically compared with exoplanet observations. Here, we review the population synthesis method and then use one population computed using the Generation III Bern model to explore how different planetary system architectures emerge and which conditions lead to their formation. The emerging systems can be classified into four main architectures: Class I of near-in situ compositionally ordered terrestrial and ice planets, Class II of migrated sub-Neptunes, Class III of mixed low-mass and giant planets, broadly similar to the Solar System, and Class IV of dynamically active giants without inner low-mass planets. These four classes exhibit distinct typical formation pathways and are characterised by certain mass scales. We find that Class I forms from the local accretion of planetesimals followed by a giant impact phase, and the final planet masses correspond to what is expected from such a scenario, the `Goldreich mass'. Class II, the migrated sub-Neptune systems form when planets reach the `equality mass' where accretion and migration timescales are comparable before the dispersal of the gas disc, but not large enough to allow for rapid gas accretion. Giant planets form when the `equality mass' allows for gas accretion to proceed while the planet are migrating, i.e., when the critical core mass is reached. The main discriminant of the four classes is the initial mass of solids in the disc, with contributions from the lifetime and mass of the gas disc. The distinction between mixed Class III systems and Class IV dynamically-active giants is in part due to the stochastic nature of dynamical interactions, such as scatterings between giant planets, rather than the initial conditions only. The breakdown of system into classes allows to better interpret the outcome of a complex model and understand which physical processes are dominant. Comparison with observations reveals differences to the actual population, pointing at limitation of theoretical understanding. For example, the overrepresentation of synthetic super Earths and sub-Neptunes in Class I systems causes these planets to be found at lower metallicities than in observations.}

\maketitle

\section{Introduction}

The advent of exoplanet discovery projects, such as the \textit{Kepler} space telescope \citep{2010ScienceBorucki}, the \textit{HARPS} survey \citep{2011MayorArxiv}, or the California Planet Search \citep{2010ApJHoward,2021ApJSRosenthal}, has led to the discovery of a large number of exoplanets. As of writing, more than \num{5000} exoplanets have been confirmed.\footnote{According to \url{http://exoplanet.eu}, the \num{5000} exoplanets mark was passed on about 27 April 2022.}

These exoplanets are diverse. Many of them have no counterpart in the Solar System. Indeed, the first discovered exoplanet around a main sequence star, 51 Peg b \citep{1995NatureMayorQueloz}, is a Jupiter-like planet orbiting its host star with a period of only \SI{4.2}{\day}. Later, a broad class of planets whose mass or size is between those of the Earth and Neptune has been unveiled \citep[e.g.][]{2017AJFulton}. The large number of discovered planetary systems permits to have statistical constraints, both on the planet properties and their correlations with other factors (such as stellar properties or stellar cluster environment).

But while we have thousands of evolved planets, there are only scant observations of planets that are still forming: the PDS 70 system \citep{2018AAKeppler,2018AAMueller,2019NatAsHaffert} and AB Aur b \citep{2022NatAsCurrie}, for instance. This limited number of observations means that we have poor direct observational constraints on the physical processes that occur during planet formation.

Parallel to the observational campaigns, there has been a wealth of works aiming at describing the physical processes that are thought to occur during the formation and later evolution of planetary systems \citep{2014PPVIBenz,2016SSRvBaruteau,2022ASSLRaymondMorbidelli}. However, the different processes cannot be tested individually because of the lack of direct data we mentioned. Nevertheless, we can leverage the large number of the end products of planetary formation (that is, the observed exoplanets) to provide indirect evidence on how these processes work. This is the basic idea of planetary population synthesis: to bridge the relative lack of observations by using the predictions of theoretical works as tests of their effectiveness at reproducing the statistical characteristics of the discovered exoplanets. This requires---among other things---a global (meaning that it includes the many interlinked physical processes occurring at each stage) and simplified (for computational requirements) formation and evolution model, which has to predict the necessary quantities for comparison with observations.

An assumption of planetary population synthesis is that the diversity of exoplanets stems from the diversity of protoplanetary discs (the initial conditions for the planetary formation process and for the numerical calculations) and the inherent chaotic nature of planetary formation (especially for \textit{N}-body interactions), rather than different models. The global model (for a pioneering work, see \citealp{2004ApJIda1}) must thus be able to reproduce the diversity of planetary systems, from terrestrial planets to giants. The interlinking of many processes included in the global model is source of complexity of the results. As such it may become advantageous to classify the final systems according to a certain scheme. In this work, we explore avenues opened by such a classification.

One focus of this work is to link the birth environment to the final planetary systems. In population synthesis, the diversity of protoplanetary discs is represented by the initial conditions. The link between environment and architecture is achieved by backtracking the initial conditions that lead to the different categories.

A global model is not only used to make predictions about the final systems; it can also be used to study many theoretical aspects of planet formation. This includes how the various processes interact with each other, which one is dominant in the different stages, or for mechanisms leading to the different categories of systems. This will be the second focus of this work. Here, we use formation pathways and analytical mass scales and match them with the final planet masses obtained in the numerical simulations for the different classes to determine which process is responsible for shaping the final planetary systems.

The final aim of planetary population synthesis is to adapt the model so that the correspondence between synthetic and observed systems is improved. To achieve this, it is necessary to understand where the model must be corrected. We apply this to the established correlations with stellar metallicity and discuss pathways for how the model can modified to improve fidelity.

This work is divided as follows. We start in Sect.~\ref{sec:principles} by discussing in more details the concepts of planetary population synthesis followed by the history and different models in Sect.~\ref{sec:history}. The remainder of this work will focus on the link between the properties of the protoplanetary discs and the diversity of planetary systems and the emergence of four classes of planetary system architectures. The methodology is presented Sect.~\ref{sec:methods} and the corresponding results in Sect.~\ref{sec:results}.

\section{Principles of Planetary Population Synthesis}
\label{sec:principles}

\begin{figure}
	\centering
	\includegraphics[width=\textwidth]{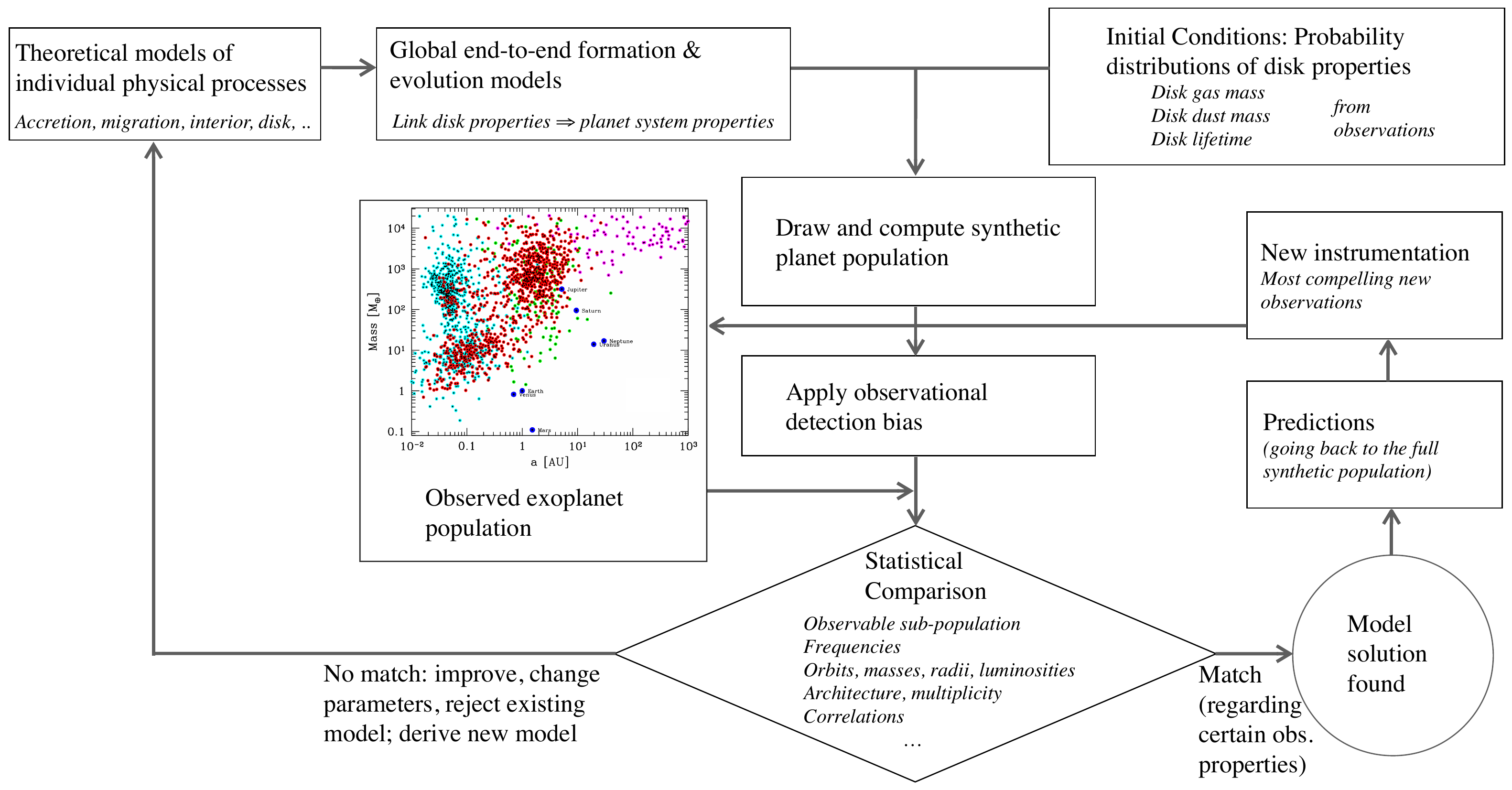}
	\caption{Principle and flow chart of planetary population synthesis method and its links to the overall theory of planet formation and to the construction of astronomical instrumentation. Extended from \citet{2015IJAsBMordasini}.}
	\label{fig:ppsprinciple}
\end{figure}

The arching goal of planetary population synthesis is to improve our understanding of planet formation by finding the theoretical model that provides the best match of the observed exoplanets from the properties of protoplanetary discs. This also enables to make predictions about the planets that have not been observed (yet). Such an endeavour encompasses many aspects, from theory to observations.

Figure \ref{fig:ppsprinciple} show how these elements are linked. The starting point are the detailed models of individual processes like disc structure and evolution, planetary solid and gas accretion, orbital migration, planet-planet interaction and so on. They inform the global end-to-end models which combine the essence of many such detailed models into one big model. This gives end-to-end models the capability to directly predict observable properties of planetary systems based on the initial conditions of the planet formation process, which are the protoplanetary disc properties.

From observations it is known that not all protoplanetary discs have the same properties, in terms of essential disc properties like mass, size, or lifetime \citep[e.g.][]{2022PPVIIManara,2022PPVIIPascucci,2022PPVIIMiotello}. Statistical disc observations allow to derive probability distributions for the initial and boundary conditions of planet formation.

Next, one can draw from these distribution and each time calculate the formation of one synthetic planetary system. By doing this many times, one obtains a synthetic planetary population. To compare quantitatively, one more step is needed: to apply the same observational detection bias that affects the actual population. Applying biases of different observational techniques on the same synthetic population makes it possible to compare different aspects of the theory, to use the full wealth of observations, and to avoid over-fitting a specific aspect.

Then, we can compare statistically and quantitatively the actual observed population with the biased synthetic population. Many different quantities can be compared, like frequencies of planet types, distributions of quantities like the mass, radius, orbital distance or the eccentricity, but also architectures of the systems, and correlations with stellar properties like stellar mass, age, and metallicity.

It is clear that given the complexity of planet formation, there will not be a perfect match between the synthetic result and the multitude of concurrently available observational constraints. These differences reveal theoretical sub-models we need to improve, where we need to change parameters or even abandon an old approach altogether, replacing it with new theoretical approaches. By this loop, the understanding of planet formation is improved.

In some cases, a model is found which is satisfactory at least in some aspect. This is of interest on the instrumental side. We can now go back to the full underlying population and study what happens if we can improve the sensitivity and capabilities of the instrumentation. These insights can be used to identify and build the next generation of astronomical instrumentation that yields the most compelling constraints for the understanding of planet formation. This then goes into the observed population, and via this second loop, planet formation theory is again improved.

There are numerous requirements to perform planetary populations synthesis. From the observational side, one needs to have extensive, statistical surveys of both protoplanetary discs and exoplanets. These survey further need to have well-characterised biases, so that models and observations can be reliably compared in a quantitative way. On the theoretical side, the model used for population synthesis must satisfy multiple conditions: 1) being able to model end-to-end formation and evolution of planetary systems, that is, from points in time of the observations of protoplanetary discs and exoplanets to ages of many gigayears, 2) predict the necessary observable quantities for comparison with exoplanet observations, and 3) having sufficient performance so that many systems can be modelled.

\subsection{Coupling between physical processes}

Conditions 1) and 2) imply that such a model must include many processes and that their coupling is also important. Major processes occurring in planetary formation (disc evolution, planetary accretion, orbital migration) all occur on similar timescales and affect each other. This fact must be reflected in models, where the processes themselves cannot be treated in isolation. This is the basis of \textit{global} models, that the physical processes must be treated concurrently and not in isolation.

\subsection{Low-dimensional approach}

The addition of condition 3) means that the model cannot treat directly first principles, i.e, solve the conservation laws in three dimensions. With the current computational capabilities, detailed, three-dimensional hydrodynamical simulations (for the protoplanetary disc, for instance) have too large requirements to be performed for the entire evolution a protoplanetary disc, let alone for an entire population of planetary systems with thousands of systems. Instead, we have to rely on more simple, lower-dimensional models.

Thus, while not using them directly, a model for planetary population synthesis relies on more sophisticated models to provide the parametrisation required for the low-dimensional approach. In turn, the simplified global models are able to test the specific models in a broader context encompassing the entire picture of planetary system formation rather than what is explorable with an isolated model, with, for instance, only orbital migration.

Another avenue for simplification of more detailed simulations is to use heuristics derived from those. This can be used for instance to mimic \textit{N}-body interactions \citep{2013ApJIda}. In the same direction, surrogate models obtained via machine learning could be used to reduce the computational requirements of more complex models \citep[e.g. as][for collisions]{2020ApJEmsenhuberA}, though this has not been yet widely used in the context of planetary population synthesis.

\subsection{Protoplanetary disc demographics: setting the stage for planetary formation}
\label{sec:pps-ppddem}

The model provides only the time evolution of planetary systems. We thus need to provide the initial conditions, which should be chosen such that they lead to the reproduction of as many disc observations as possible. The youngest observable targets are class 0 or class I discs, where the best measurements of their masses and radii come from dust emission \citep{2018ApJSTychoniec,2019ApJWilliams,2020ApJTobinA}. In contrast, measurements of class II discs \citep{2016ApJAnsdell,2016ApJPascucci} indicate that they are not suited as initial conditions as the dust content at this stage is already smaller than the estimated core masses of observed exoplanets \citep{2018AAManaraB}. Dust mass measurements are however subjects to large uncertainties, which stem from at least in part assumed temperatures and opacities \citep[e.g.][]{2022AABergezCasalou} that do not necessarily reflect the reality. The obtained values should therefore be taken with some caution.

Disc dust-to-gas ratio are usually taken to be the same as stellar metallicities \citep{2016ApJGaspar}, for which distributions have been determined from stellar spectroscopy \citep{2005AASantos,2017AJPetigura}. The distribution of disc lifetimes is estimated from measurements of the IR-excess of young stars, but the interpretation is complex \citep[coevality, stellar ages;][]{2001ApJHaisch,2009AIPCMamajek,2010AAFedele,2014AARibas,2018MNRASRichert,2021ApJMichel}.

This approach assumes that the characteristics of the  protoplanetary discs observed today are the same as those that formed the observed exoplanets. This a generally-accepted principle, but it should be kept in mind that observed exoplanets usually formed billions of years ago which is not negligible in relation to the age of the galaxy \citep{2014AAAdibekyan}. Furthermore, current disc statistics mainly stem from small, nearby star forming regions in which environmental effects differ from the birthplaces of most extrasolar planets \citep[e.g.][for more massive regions]{2020AAvanTerwisga,2022AAvanTerwisga}.

\subsection{Reproduction of exoplanets characteristics}
\label{sec:pps-exodem}

Determining which is the best model is achieved by statistically comparing to specific systems, such as the Solar System of TRAPPIST-1, and the exoplanet population as a whole. In the latter case, there are three main classes of characteristics derived from observations that synthetic populations can be compared to

First, there are planet-level statistics: their number, their distribution in observed parameters (distance, mass, radius, etc.). Hot-Jupiters (planets with a period $\lesssim\SI{10}{\day}$) are found around \num{0.5} to \SI{1}{\percent} of stars \citep{2010ScienceHoward,2011MayorArxiv}, compared to \num{10} to \SI{20}{\percent} for all giants within 5 to \SI{10}{\au} \citep{2008PASPCumming,2010PASPJohnson}. The frequency of giant planets increases with distance \citep{2013ApJDongZhu,2016AASanterne} followed by a peak in location near the iceline \citep{2019ApJFernandes,2021ApJSFulton} and a decrease beyond \num{3}--\SI{10}{\au} \citep{2016ApJBryan,2019AJNielsenA}. The planet mass function derived from radial velocity observations exhibits two peaks: one for sub-Neptunes at about \SI{10}{\mearth} and one for giants at \num{1} to \SI{2}{\mj} separated by a relative `desert' of intermediate-mass planets \citep{2011MayorArxiv}, but the significance of this `desert' is debated \citep{2021AJBennett}. Concerning planet radii, there are two peaks at \SI{1.3}{\rearth} (super Earths) and \SI{2.4}{\rearth} (sub-Neptunes) separated by a gap \citep{2017AJFulton}. Giants have a similar radius of \SI{\sim1}{\rj} leading to another local maximum \citep{2012A&AMordasiniC}.

Second, there are correlations with stellar properties. One of the first discovered is the correlation between the frequency of giant planets with stellar metallicity \citep{1997MNRASGonzalez,2004A&ASantos,2005ApJFischer}. Low-mass planets seems to be more common around low-mass stars \citep{2015ApJMuldersB} while it is the opposite for giants \citep{2013AABonfils,2018ApJGhezzi}.

Finally, we have system-level statistics. Here, we have that roughly \SI{50}{\percent} of the giant-planet-hosting systems have multiple planets \citep{2016ApJBryan}. Further multiple planets in the same system tend to have similar masses \citep{2017ApJMillholland}, radii, and orbital spacing \citep{2018AJWeiss,2021AAMishra,2022PPVIIWeiss}. This creates the interesting situation of similarity within the overall diversity.

\section{History and the different models}
\label{sec:history}

The pioneering series on extrasolar planetary population synthesis is the work of Ida and Lin. The first work, \citet{2004ApJIda1}, presented their population synthesis workflow, which contains the key ingredients that we have been discussing. They predicted the presence of a `planetary desert' because of fast growth at the onset of gas runaway (masses between \num{30} and \SI{100}{\mearth}). The second work of the series, \citet{2004ApJIda2}, reproduced the `metallicity effect' \citep{1997MNRASGonzalez,2004A&ASantos}. The third work, \citet{2005ApJIdaLin},  addressed planet formation around stars of different masses. Type~I gas driven migration became part of the model in the fourth work, \citet{2008ApJIdaLinA} and the authors found that migration rates have to be reduced by at least a factor 10 to make planet occurrence rate similar to observations. Alternatively, a `dead zone' near the iceline could reduce migration and enhance accretion so that Type~I migration does not have to be artificially reduced \citep{2008ApJIdaLinB}. The last two works of the series focused on multi-planetary systems. Finally, they added a semi-analytical approach to study planet-planet interactions without having to perform full \textit{N}-body calculation, which was applied to super-Earths \citep{2010ApJIdaLin} and giant planets \citep{2013ApJIda}. There has been some later work, for instance \citet{2018ApJIda}, who postulated that a reduced Type~II gas-driven migration is better able to match the location very massive planets (super-Jupiters) near \SI{1}{\au} while less massive planets (Saturn to Jupiter mass) being found across a wider range of distances. Also, \citet{2020MNRASMiguel} apply an updated code to the case of late M and brown dwarfs.

Another series of population synthesis works were performed by Mordasini and collaborators. The first series, based on the formation model of \citet{2004A&AAlibert,2005A&AAlibert} considered only a single embryo per disc and found that migration had to be reduced by factor between 10 and 100 to avoid planets falling into the star \citep{2009A&AMordasinia}. Outcomes were then compared to exoplanets discovered by radial velocity at the time \citep{2009A&AMordasinib}. This work was later extended to different stellar masses \citep{2011A&AAlibert}. The last work of that series, \citep{2012A&AMordasiniA}, focused on the disc properties required to form giant planets. The later work took to directions, with one branch focusing on predicting planet radii and luminosities for comparison with transit and direct imaging surveys \citep{2012A&AMordasiniB,2012A&AMordasiniC}. The other branch focused on the formation of multi-planetary systems \citep{2013A&AAlibert}, better representation of solids accretion \citep{2013A&AFortier}, and the planets' chemical composition \citep{2014AAThiabaud,2015A&AThiabaud}. These two branches were later merged to compute multi-planet populations capable of predicting the different properties for various observational techniques for both Solar-mass stars \citep{2021AAEmsenhuberA,2021AAEmsenhuberB} and M-dwarfs \citep{2021AABurn}. Related works include the correlation between close-in super Earths and distant giant planets \citep{2021AASchleckerA}, the reproduction of the formation model model using machine learning techniques \citep{2021AASchleckerB}, and the reproduction of peas-in-a-pod structure and comparison with Kepler \citep{2021AAMishra}.

A series of population synthesis works by Alessi and Pudritz focuses on the effect of migration traps, disc properties, and chemistry. The first work \citep{2018MNRASAlessi} focuses on the conditions needed to obtain a separation between hot and warm Jupiters in terms of dead zone and envelope metallicity. Then, the authors study the effects of the discs initial properties \citep{2020MNRASAlessiA} and the predictions on core and envelope composition \citep{2020MNRASAlessiB}. Their latest work focuses on the inclusion of magnetised disc winds as transport mechanism \citep{2022MNRASAlessiPudritz}.

There are also studies based on the core accretion paradigm assuming that the solids are accreted in the form of pebbles, but comparatively less than with planetesimals. One is \citet{2018ApJChambers}, who finds that low-viscosity and other parameters that provide for short accretion timescales are best able to match exoplanets observations. Another work by \citet{2020AALiu} focus on low-mass stars with a single-embryo model going from M-stars down to brown dwarfs. With multiple interacting bodies, \citet{2019AALambrechts} focus on the assembly of terrestrial or super-earth systems while \citet{2019AABitschA} benchmark their models against the giant planet population. There is also a series of work focusing on the stellar environment: \citet{2018MNRASNdugu} investigated how the background heating by nearby stars, while \citet{2022MNRASNdugu} investigated how stellar encounters affect the formation, for instance by disc truncation. A comparison of pebble and planetesimal based models in a population synthesis work was done in \citet{2020AABrugger}.

Two recent Protostars and Planets VII review chapters address planetary population synthesis. \citet{2022PPVIIWeiss} investigates compact multi-planetary systems while \citet{2022PPVIIDrazkowska} focuses on the comparison between planetesimals and pebble accretion.

The base assumption that all extrasolar planets have formed by a bottom-up formation pathway (i.e. by some variant of the core accretion paradigm) might not hold \citep{2019ScienceMorales,2022NatAsCurrie}. Therefore, several population synthesis models were also built based on the gravitational instability paradigm \citep{2013MNRASForganRice,2018MNRASForgan,2015MNRASNayakshinB,2018ApJMueller}. These models simulate the evolution of the gas disc, use fragmentation criteria, and follow the evolution of the emerging gas clumps via tidal downsizing, migration, and contraction.

\section{Methods}
\label{sec:methods}

In this work, we use the Generation~III \textit{Bern} model of planetary formation and evolution, as it was presented in earlier works \citep{2021AAEmsenhuberA}. Thus, we only provide a summary of the model here. The interested reader is referred to relevant works for an in-depth description of the model and the population synthesis methods \citep{2021AAEmsenhuberA,2021AAEmsenhuberB,2021AABurn}.

\begin{figure}
	\centering
	\includegraphics[width=\textwidth]{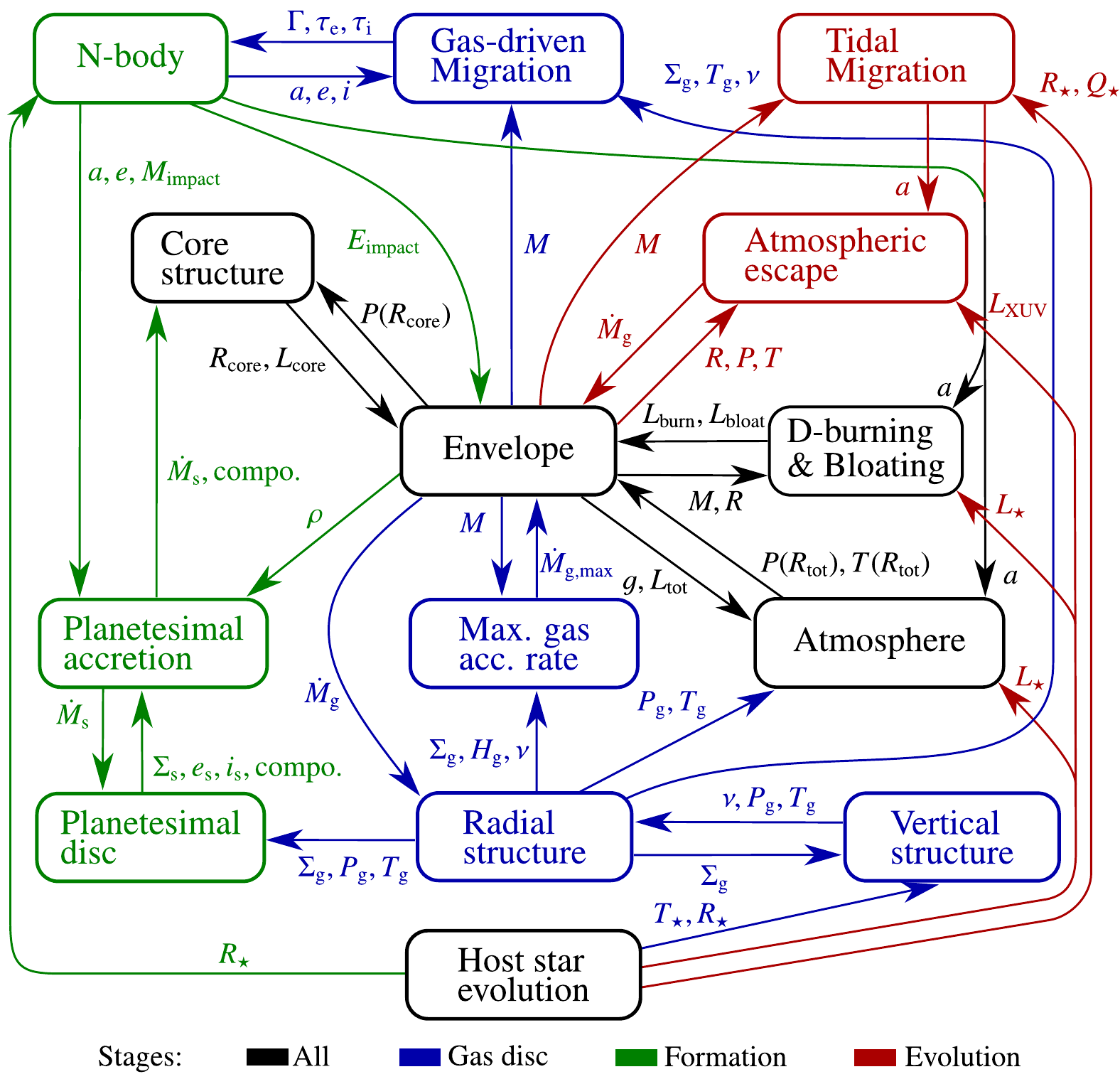}
	\caption{Sub-models in the Generation III Bern Model of planet formation and evolution. It is an example of a comprehensive end-to-end model as often used in planetary population synthesis calculations. Arrows indicate the most important exchanged quantities. Colors give the phase when a  sub-module is active. For computational time constraints, the `Formation' phase is assumed to  last for \SI{100}{\mega\year} only (extended from the \SI{20}{\mega\year} of \citealp{2021AAEmsenhuberB}), because simulating with an N-body integrator thousands of planetary system over billion year-timescales is currently prohibitively expensive. In the subsequent `Evolution' phase, the planets evolve individually to a \SI{5}{\giga\year} age. Updated from \citet{2021AAEmsenhuberA}.}
	\label{fig:modelscheme}
\end{figure}

The \textit{Bern} model \citep{2004A&AAlibert,2005A&AAlibert,2013A&AAlibert,2009A&AMordasinia,2012A&AMordasiniB} couples the formation and evolution stages. The formation stage follows the evolution of a viscously-accreting protoplanetary disc, the dynamical state of planetesimals in the disc, concurrent accretion of solids and gas from the disc by the protoplanets, gas-driven planetary migration, and dynamical (N-body) interactions between multiple protoplanets in a each system. The evolution stage tracks the thermodynamic evolution (cooling and contraction), atmospheric escape, and tidal migration of each planet individually. Figure \ref{fig:modelscheme} shows the most important sub-modules of the model and the key quantities that they are exchanging (the model is entirely described in \citealp{2021AAEmsenhuberA}). The individual sub-modules can be of considerable complexity, but numerically use---with the exception of the N-body integrator---a low-dimensional approach (for example the planetary interiors are assumed to be 1D spherically symmetric). This approximation must be made to keep the ability to simulate the planetary systems from their origins to the present day, which is only possible with low-dimensional models because of computational time constraints.

\subsection{Formation stage}

The gas component of the protoplanetary disc is modelled with a 1D radial, axisymmetric representation that is viscously accreting \citep{1974NMRASLyndenBellPringle}. Viscosity is computed following the standard $\alpha$ parametrisation \citep{1973A&AShakuraSunyaev} with a fixed value of $\alpha=\num{2e-3}$, while the vertical structure is computed in a vertically-integrated approach \citep{1994ApJNakamoto}. We include additional sink terms to the evolution to account for internal and external photoevaporation, following \citet{2001MNRASClarke} and \citet{2003ApJMatsuyama}, respectively, and the accretion by the protoplanets. The solids in the disc are assumed to be in the form of planetesimals of a fixed radius $r=\SI{300}{\meter}$. They are treated as a 1D radial, axisymmetric surface density (similar to the gas) with a dynamical state (eccentricity and inclination). The dynamical state is evolved under the stirring by the other planetesimals, the protoplanets, and the damping due to the presence of a gas disc \citep{2002IcarusOhtsuki,2006IcarusChambers,2013A&AFortier}.

Planet formation follows the core accretion paradigm \citep{1974IcarusPerriCameron,1978PThPhMizuno,1980PThPhMizuno}. Initially, a given number of embryos are randomly placed in the disc with a uniform spacing in the logarithm of the distance, so that they are separated by a given amount of Hill radii, as found in \textit{N}-body simulations of embryo formation \citep{2002ApJKokubo}. They will accrete planetesimals, which is taken to be in the oligarchic regime \citep{1993IcarusIdaMakino,2001IcarusInaba,2003IcarusThommes}. The gas envelope structure is initially in contact with the surrounding disc and accretion occurs thanks to radiating away the gravitational binding energy released during this process, i.e., on a Kelvin-Helmholtz timescale \citep{1996IcarusPollack,2015ApJLeeChiang}. The gas accretion rate is thus determined by solving the 1D spherically symmetric internal structure equation of the envelope \citep{1986IcarusBodenheimerPollack}. The presence of a gaseous envelope also enhances the capture radius of planetesimals, which we compute following the procedure of \citet{2003A&AInaba}. As a protoplanet core grows, its Kelvin-Helmholtz timescale decreases \citep{2000ApJIkoma}. The gas accretion rate thus increases until it becomes larger than what can be supplied by the disc. When this occurs, the envelope can no longer maintain the contact with the disc and contracts \citep{2000IcarusBodenheimer}. In this stage, the structure equations are used to compute the planet radius instead, while the gas accretion is set by the supply from the disc \citep{2012A&AMordasiniB,2012A&AMordasiniC}.

Dynamical interactions between the different protoplanets are modelled by means of the \texttt{mercury} \textit{N}-body package \citep{1999MNRASChambers}. Also, type~I gas-driven migration is computed according to \citet{2014MNRASColemanNelson} while type~II and the transition between the two regimes follow \citet{2014A&ADittkrist}. Migration is applied as additional forces in the \textit{N}-body.

\subsection{Evolution stage}

After a fixed time of \SI{100}{\mega\year} (compared to \SI{20}{\mega\year} in \citealp{2021AAEmsenhuberB} to increase the duration of the \textit{N-body} interactions), the model transitions to the evolution stage, where the planet are individually evolved to \SI{5}{\giga\year}. In addition to the normal thermodynamic evolution (cooling and contraction) of the interior which is found by solving 1D internal structure equations, this stage also tracks XUV-driven atmospheric escape \citep{2014ApJJin,2018ApJJin} and inward migration due to stellar tides  \citep{2008CeMDAFerrazMello,2011AABenitezLlambay}.

\subsection{Analysis: mass scales}
\label{sec:timescales}

Because of the numerous interaction and feedback (see the many arrows in Fig.~\ref{fig:modelscheme}), global models can produce numerical results of considerable complexity which can be difficult to interpret. To understand what leads to the final shape of the final systems, it is therefore helpful to analytically investigate a number of mass scales of different processes that take place during the formation stage. The purpose is to compare these analytical mass scales to the numerical formation tracks to understand which process is linked to the different features.

The first two mass scales are related to the accretion of planetesimals only. Temporally, the first process that occurs is this accretion of planetesimals. Absent radial motion caused by the gas disc (orbital migration) or interactions with other bodies, the embryos can only accrete from nearby planetesimals, which will halt once the local reservoir is depleted. This leads to the planetesimal isolation mass \citep{1987IcarusLissauer},
\begin{equation}
    M_\mathrm{iso}=\frac{\left(2\pi b a^2\Sigma_\mathrm{p}\right)^\frac{3}{2}}{\sqrt{3M_\star}} \sim \SI{3.6e-2}{\mearth}\left(\frac{a}{\SI{1}{\au}}\right)^3\left(\frac{\Sigma_\mathrm{p}}{\SI{10}{\gram\per\square\centi\meter}}\right)^\frac{3}{2}\left(\frac{M_\star}{\SI{1}{\msun}}\right)^{-\frac{1}{2}},
    \label{eq:miso}
\end{equation}
with $b\sim10$ the full width of the feeding zone in Hill radii of the planet $r_\mathrm{H}=a\left(M/(3 M_\star)\right)^{1/3}$, $a$ the local position, $\Sigma_\mathrm{p}$ the local surface density of planetesimals, and $M_\star$ the mass of the central star. At large separation, however, the accretion timescales become so slow that the isolation mass cannot be reached in any reasonable time. There, growth is rather stalled at a time comparable to the dispersal of the protoplanetary disc, which halts the damping of the planetesimals random velocities and causes in turn the accretion rate to strongly decrease. We can use the analytic accretion timescales in the oligarchic regime \citep{1993IcarusIdaMakino} to estimate a growth mass \citep[see][Eq. 14 for the derivation]{2003IcarusThommes},
\begin{eqnarray}
    \label{eq:mgrow}
    M_\mathrm{grow} &\sim& \dot{M}\tau_\mathrm{acc}= k_\mathrm{grow}\frac{t_\mathrm{disc}^3 M_\star^{1/2} \Sigma_\mathrm{g}^{6/5} \Sigma_\mathrm{p}^3}{\rho_m^{4/5} \rho_M a^{3/2} m^{2/5} (H/a)^{6/5}} \\
    &\sim& \SI{6.8}{\mearth}\left(\frac{t_\mathrm{disc}}{\SI{1e6}{\year}}\right)^3\left(\frac{\Sigma_\mathrm{p}}{\SI{10}{\gram\per\square\centi\meter}}\right)^3\left(\frac{a}{\SI{1}{\au}}\right)^{-\frac{3}{2}}\left(\frac{M_\star}{\SI{1}{\msun}}\right)^\frac{1}{2} \left(\frac{\Sigma_\mathrm{g}}{\SI{2400}{\gram\per\centi\meter\squared}}\right)^{6/5}\nonumber\\
    && \left(\frac{\rho_m}{\SI{1}{\gram\per\centi\meter\cubed}}\right)^{-4/5} \left(\frac{\rho_M}{\SI{3.2}{\gram\per\centi\meter\cubed}}\right)^{-1} \left(\frac{m}{\SI{1e18}{\gram}}\right)^{-1/10} \left(\frac{H/a}{0.05}\right)^{-6/5}
\end{eqnarray}
with a constant prefactor $k_\mathrm{grow} = \frac{ \pi^3 3^{21/15}}{4^{3/5} } \left(\frac{C_D b}{C_e}\right)^{6/5} G^{3/2} \sim 11.9 \sqrt{G^3}$. Here, $G$ is the Newtonian gravitational constant and reasonable values are for the drag coefficient $C_D\sim 1$ \citep{1976PThPhAdachi}, for the viscous stirring factor $C_e \sim 40$ \citep{1993IcarusIdaMakino,2002IcarusOhtsuki}, and for the width of the feeding zone $b\sim 10$. Further, $\Sigma_\mathrm{g}$ is the local gas surface density, $H$ the disc scale height often combined to the aspect ratio $H/a \sim 0.1$, $t_\mathrm{disc}$ the lifetime of the gaseous disc, $\rho_{m}$ and $\rho_{M}$ are the bulk densities of the planetesimals and the protoplanet, and $m$ is the mass of one planetesimal. Note that while $M_\mathrm{grow}$ is much larger than the other scales at short distances (within \SI{1}{\au}), it decreases most rapidly with distance for realistic disc and planetesimal surface density profiles. It should be noted that $M_\mathrm{grow}$ only makes sense in the outer parts of the disc (outside of $\sim10$ AU) where it holds that $M_\mathrm{grow}\leq M_\mathrm{iso}$.

The next two mass scales are related to orbital migration. As the mass of a protoplanet increases, its oligarchic planetesimal accretion rate decreases, whereas its type I migration rate increases. At one point migration becomes faster than growth. In a mass-distance diagram this means that planetary formation track bend from a vertical (upward) direction inward and become more horizontal. This happens at what we call the `equality mass', where the migration timescale is equal to the accretion timescale. It can be estimated by equating the planetesimals accretion timescale from (\citealp{2018BookMordasini}, Eq. 10, or equivalently \citealp{2003IcarusThommes}) with the type~I migration timescale neglecting non-linear corotation torques (see \citealp{2002ApJTanaka}, Eq. 70, but with non-isothermal torques from \citealp{2010MNRASPaardekooper}). This results in
\begin{equation}
    M_\mathrm{eq} = k_\mathrm{eq}\frac{\Sigma_\mathrm{p}^{3/4} M_\star^{5/4} (H/a)^{6/5}}{\Upsilon^{3/4}\Sigma_\mathrm{g}^{9/20} \rho_{m}^{1/5} \rho_M^{1/4} a^{3/4} m^{1/10}}\,,
    \label{eq:meq}
\end{equation}
where $k_\mathrm{eq} \simeq 3^{11/10} \pi^{3/4} \left(\frac{b C_D}{2 C_e}\right)^{3/10} \simeq 4.2$, $\Upsilon \simeq \gamma^{-1}(2.5 +1.7 \beta_\mathrm{T} - 0.1 \beta_\Sigma$) is an order unity prefactor for the Lindblad torque which depends on the adiabatic index of the gas $\gamma$, the radial slopes of the surface density set to $-\beta_\Sigma$, and the one of temperature to $-\beta_\mathrm{T}$ \citep{2010MNRASPaardekooper} and $m$ is the mass of an individual planetesimal. Entering typical values, this becomes
\begin{eqnarray}
    M_{\mathrm{eq}} &= \SI{1.2}{\mearth} &\left(\frac{\Upsilon}{2.23}\right)^{-3/4} \left(\frac{M_{\star}}{\SI{}{M_\odot}}\right)^{5/4} \left(\frac{\Sigma_\mathrm{p}}{\SI{10}{\gram\per\square\centi\meter}}\right)^{3/4} \left(\frac{m}{\SI{e18}{\gram}}\right)^{-0.1}\\
    &&\left( \frac{H/a}{0.05}\right)^{6/5} \left(\frac{\Sigma_\mathrm{g}}{\SI{2400}{\gram\per\square\centi\meter}}\right)^{-0.45}  \left(\frac{a}{\SI{1}{\au}}\right)^{-3/4} \left(\frac{\rho_m}{\SI{1}{\gram\per\centi\meter\cubed}}\right)^{-1/5} \left(\frac{\rho_M}{\SI{3.2}{\gram\per\centi\meter\cubed}}\right)^{-1/4}.
\end{eqnarray}

However, the classical type I Lindblad torques can be counteracted by the co-rotation torque \citep{1991LPIWard,2001ApJMasset}. The co-rotation region exerts a torque on the planet if gradients in specific vorticity and entropy are present. To not balance out those gradients and saturate the corotation torque, the gradients need to be restored by the disc's viscosity faster than the time it takes for a parcel of gas to perform a full horseshoe orbit \citep{2011MNRASPaardekooper,2014A&ADittkrist}. The width of the corotation region is given by
\begin{equation}
    \Delta a_{\rm HS} = C_{\rm HS} \frac{a}{\gamma^{1/4}} \sqrt{\frac{M a}{M_\star H}},
\end{equation}
where $C_{\rm HS} \approx 1.1$ is a numerical factor determined from numerical experiments \citep{2010MNRASPaardekooper} and $\gamma$ is the heat capacity ratio or adiabatic index. Since $\Delta a_{\rm HS}$ increases with planetary mass, we can determine a critical mass at which the corotation torque saturates by equating the libration and the viscous diffusion timescale across the corotation region \citep{2012MNRASHellaryNelson}
\begin{equation}
    \frac{8 \pi a}{3 \Omega_\mathrm{K} \Delta a_{\rm HS}} \stackrel{}{=} \frac{(\Delta a_{\rm HS})^2}{\nu}.
\end{equation}
Solving for the mass results in what we call the saturation mass
\begin{equation}
     M_\mathrm{sat} = \left(\frac{8 \pi \alpha }{3}\right)^{2/3} \sqrt{\frac{\gamma}{C_{\rm HS}^4}} M_\star \left(\frac{H}{a}\right)^{7/3}.
     \label{eq:msat}
\end{equation}
This mass scale the upper limit where planets can be trapped between outward and inward migration regions (migration traps). Above this value, planets always migrate towards the star. However, planets below this mass scale are not necessarily trapped into convergence zones, as that depends on the relative strength of the Lindblad and corotation torques. We note that this mass scale depends on $H/a$, which is distant- and time-dependent. This leads to time-dependent predictions, as the aspect ratio decreases as a function of time because of decreasing viscous heating, and stellar evolution \citep[e.g.][]{2015AABaraffe}. The assumption of viscous dissipation as the radial transport method in the disc does also play a major role here, as the inclusion of magnetically-driven disc winds combined a lower viscous dissipation would severely reduce the corotation torques (though it may also produce large outward corotation torques under certain conditions, \citealp{2020AAKimmig}).

The fifth mass scale is linked to giant impacts. This important mass scale does not directly depend on orbital migration but may depend indirectly on it, as migration leads to a radial redistribution of the protoplanets: the final stage of the inner planets' formation is dominated by giant impacts. This phase can lead to planet masses significantly larger than the planetesimal isolation mass. Assuming that the damping of the protoplanets' eccentricities by remaining planetesimals is weak, the resulting mass can be estimated \citep{2004ApJGoldreich}. To obtain this mass scale, we assume that the random velocity are given by mutual gravitational interactions between the planets only. Therefore, the derivation of this mass scale is similar to the isolation mass, but the extent of the feeding zone is modified to include the eccentric excursions from apastron to periastron of the growing planets, and the building blocks are no longer directly the planetesimals, but larger protoplanets with a surface density $\Sigma_\mathrm{P}$. The captial `P' here stands in contrast to that of the planetesimals $\Sigma_\mathrm{p}$. Here, we assumed for an order-of-magnitude estimate that, locally, all planetesimals have been accreted by the protoplanets, for example bodies that have reached the local planetesimal isolation mass. In detail, this assumption does not hold in regard to the Solar System where there are planetesimal-like bodies that remain until today, but we are here interested at the overall mass budget, which is dominated by the planets. This assumption is also justified a posteriori from the fact that all planetesimals are accreted in the terrestrial region (Sect.~\ref{sec:res-linkiniconds}). The radial distribution of these protoplanets may differ from the initial planetesimals depending on whether orbital migration lead to a significant redistribution. Let $m$ be the mass of the planet which is given by
\begin{equation}
m = 2 \pi a 2 \Delta a \Sigma_\mathrm{P}
\end{equation}
for a radial extent of the feeding zone $\Delta a = ea$ if the orbit is eccentric enough such that the width used for the isolation mass (Eq.~(\ref{eq:miso})) is much smaller than the radial excursion of the planet, that is when $\Delta a\gg r_\mathrm{H}$ or $e\gg\left(M/(3M_\star)\right)^{1/3}$. Given multiple interacting bodies, the planets' random velocity is approximately given by the planets' escape velocity $v_{\mathrm{esc}}=\sqrt{2 G m/R}$ which then leads to an eccentricity of $e \approx v_{\mathrm{esc}}/v_{\mathrm{K}}$. Thus,
\begin{equation}
m = 4 \pi a^2 \Sigma_\mathrm{P} \sqrt{\frac{2 G m / R}{G M_\star/a}}\,,
\end{equation}
where we can assume a constant bulk density $\rho_\mathrm{p}$ for each planet and solve for $m$ to get the Goldreich mass
\begin{equation}
M_{\mathrm{Gold}} = 16 a^3  \Sigma_\mathrm{P}^{3/2} \left(\frac{2\pi^7 a^3 \rho_\mathrm{P}}{3 M_\star^3}\right)^{1/4}\,.
\label{eq:mgold}
\end{equation}
If the effect of a previous orbital migration of the protoplanets can be neglected, we can set $\Sigma_\mathrm{P}=\Sigma_\mathrm{p}$. It should be noted that the Goldreich mass is inherently a mass scale that is not strictly local, as it is obtained via a series of giant impacts. This leads to a radial mass exchange, as can be seen in the simulations below. It is therefore useful to calculate $M_{\mathrm{Gold}}$ not with local quantities only, but take an average over the region where the giant impact phase takes place (like for example inside of the iceline, as it is done in the next section).

Finally, two additional mass scales are important for giant planet formation: the critical (core) mass for gas runaway accretion \citep{1978PThPhMizuno,1980PThPhMizuno,2014ApJPisoYoudin} and the mass where planets switch from faster type I to slower type II migration \citep{2006IcarusCrida}. Regarding the former, it is important to directly calculate it with the structure equations, as semi-analytical expression can give grossly incorrect results \citep{2019AAAlibertVenturini}.

\section{Results}
\label{sec:results}

We conduct our main analysis for a synthetic planet population that is based on the nominal population that  was presented in \cite{2021AAEmsenhuberB} (\texttt{NG76}). However, in the simulation  shown here (\texttt{NG76longshot}), the formation stage which includes \textit{N}-body interactions was extended from 20 to \SI{100}{\mega\year}. As a simplification, the model starts with all the solids are in the form of planetesimals, with a steeper radial slope than the gas to account for planetesimal formation simulations (power-law index of $\beta_\mathrm{s}=-1.5$ following \citealp{2020AAVoelkel}). The solid component is thus always in the form of planetesimals and does not account for the presence of dust in evolved discs. At the same time, \num{100} lunar-mass embryos are placed in each of the \num{1000} simulated planetary systems. This population contains a variety of planetary systems, from terrestrial planet systems to multiple giants, as we will show in this section.

\subsection{Four classes of system architectures}

To analyse the formation patterns of different kinds of planetary systems and determine how the diversity of protoplanetary discs and environment affect the formation, we start by classifying the final systems into a number of classes of system that show similar properties. We base the classification on mainly the mass-distance diagram plus additionally the bulk composition. For the latter, we distinguish five compositions: Jovian (H/He dominated in mass), Neptunian (volatile (ice)-rich with H/He), water worlds (volatile-rich without H/He), hyterran (silicate-iron core with H/He envelope) and terrestrial (silicate-iron without H/He or ices).

By visually inspecting each of the \num{1000} synthetic planetary systems in the population, we find that the system architectures emerging in the simulations can be divided into four main classes. These are, ordered in increasing typical mass of formed planets, 1) (compositionally) ordered Earths and ice worlds systems, 2) migrated sub-Neptune systems, 3) mixed systems with low-mass and giant planets, broadly speaking akin to the Solar System, and 4) dynamical active giants. The preponderance of the different classes is as follows: \SI{57.6}{\percent} are part of Class I, \SI{22.1}{\percent} of the system are found to be part of Class II, \SI{12.3}{\percent} are part of Class III, and finally \SI{8.0}{\percent} are found to be part of Class IV. As was already discussed by \citet{2021AAEmsenhuberB}, most systems contain only low-mass planets (Class~I and~II) while only about \SI{20.3}{\percent} of the systems are in classes that contain giant planets (Class~III and~IV). However, these numbers are sensitive to the choice of initial conditions, which themselves are subject to uncertainties (Sect.~\ref{sec:pps-ppddem}).

We find that the extension of the formation stage to \SI{100}{\mega\year} (compared to \SI{20}{\mega\year} in \citealp{2021AAEmsenhuberB}) affects mostly the Class-I systems. The final architecture of the systems in the other classes is only seldom affected by this change. In the following subsections, we will review the formation pathways of each of the four classes and provide example of system architecture in each of them.

Our classification of planetary systems is based on a human classificator. As becomes clear below, the classes found this way correspond to fundamentally different temporal formation pathways which are governed by different physical processes (like giant impacts, orbital migration, or gas runaway accretion). These processes in turn manifest themselves through the different mass scales introduced in Sect.~\ref{sec:timescales}. Our approach thus stresses the temporal emergence of different system architectures and the (theoretical) physical mechanisms included in the model leading to them. A distinct approach was recently presented by \citet{2023AAMishraA,2023AAMishraB} who also classified the population used here in four distinct types. Their analysis is focused on observable systems, by applying observational biases on the final systems. They also employed a number of coefficients that can be calculated numerically out of the physical properties to identify the classes instead of a human classificator. This approach uses the current state of the systems and is more  observationally-driven, in contrast to the temporal/evolutionary aspect and the underlying theoretical physical processes that are of import in the approach used here. Interestingly enough, there is nevertheless a significant degree of convergence in the outcomes of the two approaches, as discussed in Sect.~\ref{sec:architectureclass}.

\subsubsection{Class I: Compositionally ordered Earths and ice worlds systems}

\begin{figure}
	\centering
	\includegraphics[width=\textwidth,height=17cm]{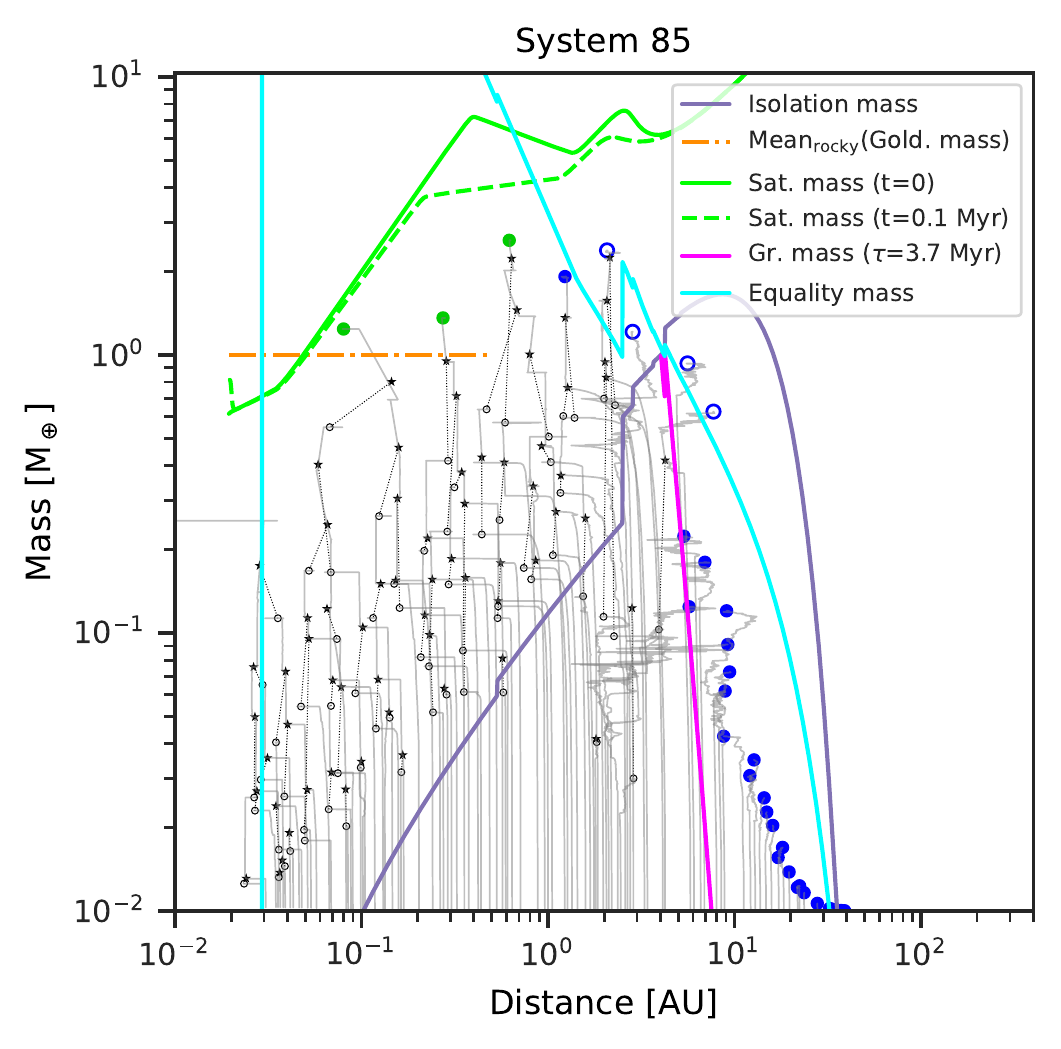}
	\caption{One example of formation tracks for a system that belongs to Class I. The solid grey lines show the tracks of all protoplanets. The ones that remain through the end (\SI{5}{\giga\year}) are additionally marked with colour symbols at the end of the track, according to their composition. Green (blue) full circles show rocky (icy) bodies without H/He, while green (blue) empty circles show rocky (icy) bodies with a H/He envelope, and red full circles are gas-rich bodies (more than \SI{50}{\percent} H/He by mass). Black stars and empty circles show target and impactor of giant impacts, with dotted black lines connecting the impact partners. The analytical isolation (violet, Eq.~(\ref{eq:miso})), Goldreich (orange, Eq.~(\ref{eq:mgold})), saturation (green, Eq.~(\ref{eq:msat})), equality (cyan, Eq.~(\ref{eq:meq})), and growth (pink, Eq.~(\ref{eq:mgrow})) mass scales are shown. The latter is only shown where it is lower than isolation mass. In Class I systems, the masses in the inner system are governed by the (mean) Goldreich's mass $M_\mathrm{Gold}$, i.e., via the final giant impact phase. The saturation mass is evaluated at two different times.}
	\label{fig:class1}
\end{figure}

\begin{figure}
	\centering
	\includegraphics[width=\textwidth]{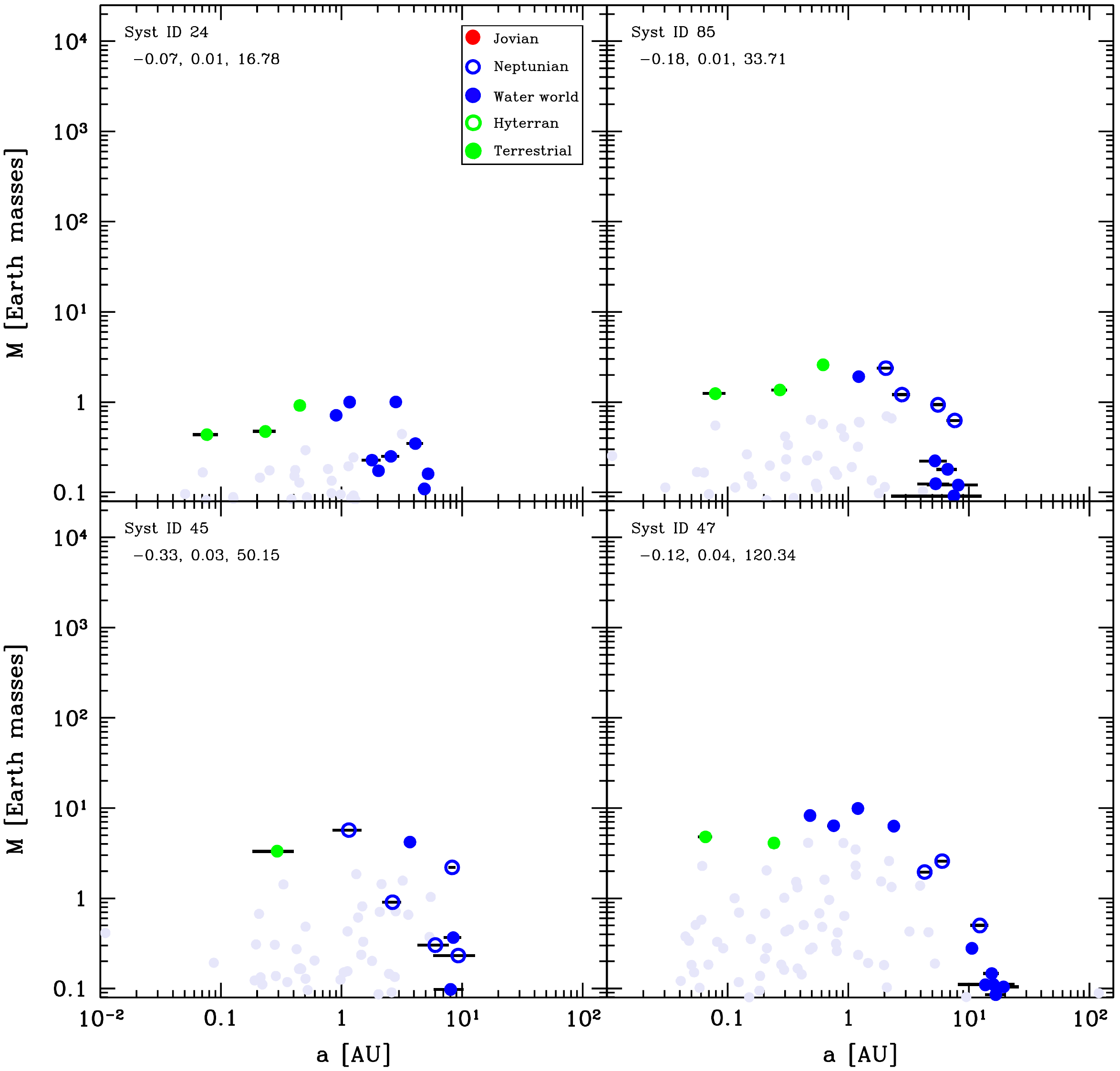}
	\caption{Final architecture of four example systems in Class I. Coloured dots show the position in the mass-semimajor axis plane of different planet composition types as indicated in the legend. The three numbers in the top left give the initial conditions [M/H], initial gas disc mass [\si{\msun}], and initial mass of planetesimals [\si{\mearth}]. The systems are ordered according to the latter. Horizontal black bars indicate the periastron to apastron distance, reflecting  orbital eccentricity. Grey full circles show the last position of protoplanets that were lost. The ones lost by accretion onto the star are put on the left y-axis of each panel and those ejected on the right one. The remaining grey points correspond to protoplanets that were accreted by another more massive protoplanet. The point then shows the mass and semi-major axis just before this accretion event.}
	\label{fig:class1examples}
\end{figure}

Class I contains systems of low-mass planets whose composition remain well-ordered in the sense that there are rocky (silicate/iron) planets inside and icy (volatile-rich) planets outside. One example of the formation tracks of such systems is provided in Fig.~\ref{fig:class1}. For the embryos that start inside the (water) ice line, growth is divided in three phases. The first phase is in-situ accretion of planetesimals, until the planets reach the planetesimal isolation mass. As the isolation mass is much lower than the equality mass in this area, migration is negligible. Only the few most massive planets, those just beyond the iceline, become massive enough to undergo some migration. The second phase is some inward migration without accretion. Migration is due to the push by the more massive outer planets, meaning that large resonant convoys form. The tight spacing of the embryos combined with the short accretion timescales mean that all the planetesimals inside the iceline have been accreted, rendering further solid growth impossible. The planets are also too low mass for gas accretion. In addition, as the planetesimal accretion timescale increases with distances, the inner planets have finished growth by planetesimals by the time the outer planets begin to migrate, leading to migration at constant mass. The third phase, which starts when the gas disc vanishes, is growth by giant impacts, which allows the planets to reach masses that are nearly two orders of magnitude larger than the isolation mass, as was discussed by \citet{2021AAEmsenhuberA}. The masses are not large enough to either be substantially affected by migration (when the gas disc is still present) or for system-wide destabilisation during the giant impact phase. Thus, the distinction between rocky and icy planets remains, as material is not reshuffled in the system; the only effect is an overall shift to closer distance, which lead to the presence of icy planets down to about \SI{1}{\au}. This is inside by about 1 -- \SI{2}{\au} of the location of the iceline (around \SI{3}{\au} in these systems; its location is set from the initial disc profile and then kept constant). In the inner system where many giant impacts have happened, the final planet masses are comparable to, or somewhat larger than the Goldreich mass. Its mean value (mean over the region inside of the iceline) is shown in the figure. In the outer region, there is a sharp decrease of the planet masses beyond the location where the growth timescale is larger than the disc lifetime. It should be noted that we have simulated the formation phase which in particular includes the accretion of planetesimals and the \textit{N}-body interaction including giant impacts for \SI{100}{\mega\year}. Further growth may occur over very long timescales \citep{2004ApJIda1}. The extension from \SI{20}{\mega\year} in \citet{2021AAEmsenhuberB} allows for continued growth of the terrestrial planets by giant impacts. Class-I systems presented here have therefore on average less terrestrial planets (but more massive) than those in the original population from \citet{2021AAEmsenhuberB}. However, only a small fraction of systems would have their architecture affected in such a way that this would alter their classification.

Four examples of final Class~I architectures are shown in Fig.~\ref{fig:class1examples}. Here we see that the systems exhibit a similar behaviour. The inner planets tend to have slightly increasing mass with distance. This follows the behaviour of the planetesimal isolation mass, which for the assumed MMSN-like initial planetesimal surface density profile $\propto a^{-1.5}$ increases with distance. However, the mass increase is not as strong as the radial dependency of the isolation mass that is shown with the purple line in Fig.~\ref{fig:class1}. This implies that the giant impact phase is able to reduce the effect of the planetesimals isolation mass; rather the Goldreich mass is the relevant scaling for the final masses at the end of the giant impact phase. One trend that can be observed from the systems shown here is that there is an anticorrelation between the number of planets and their masses. The systems in the top two panels each have three terrestrial planets whose masses are \SI{\sim1}{\mearth} while the systems in the two bottom panels have one or two terrestrial planets, whose masses are \SI{\sim4}{\mearth}. We finally observe the peas-in-a-pod effect \citep{2021AAMishra,2022PPVIIWeiss} across the entire population, where planets in a system have similar masses. The characteristic mass scale, the Goldreich mass, depends on the solid building block surface density (Eq.~(\ref{eq:mgold})). Therefore, the higher the initial mass of planetesimals in a disc (initial mass in \si{\mearth} given in the figure), the higher the Goldreich mass.

\subsubsection{Class II: Migrated sub-Neptunes systems}

\begin{figure}
	\centering
	\includegraphics[width=\textwidth,height=17cm]{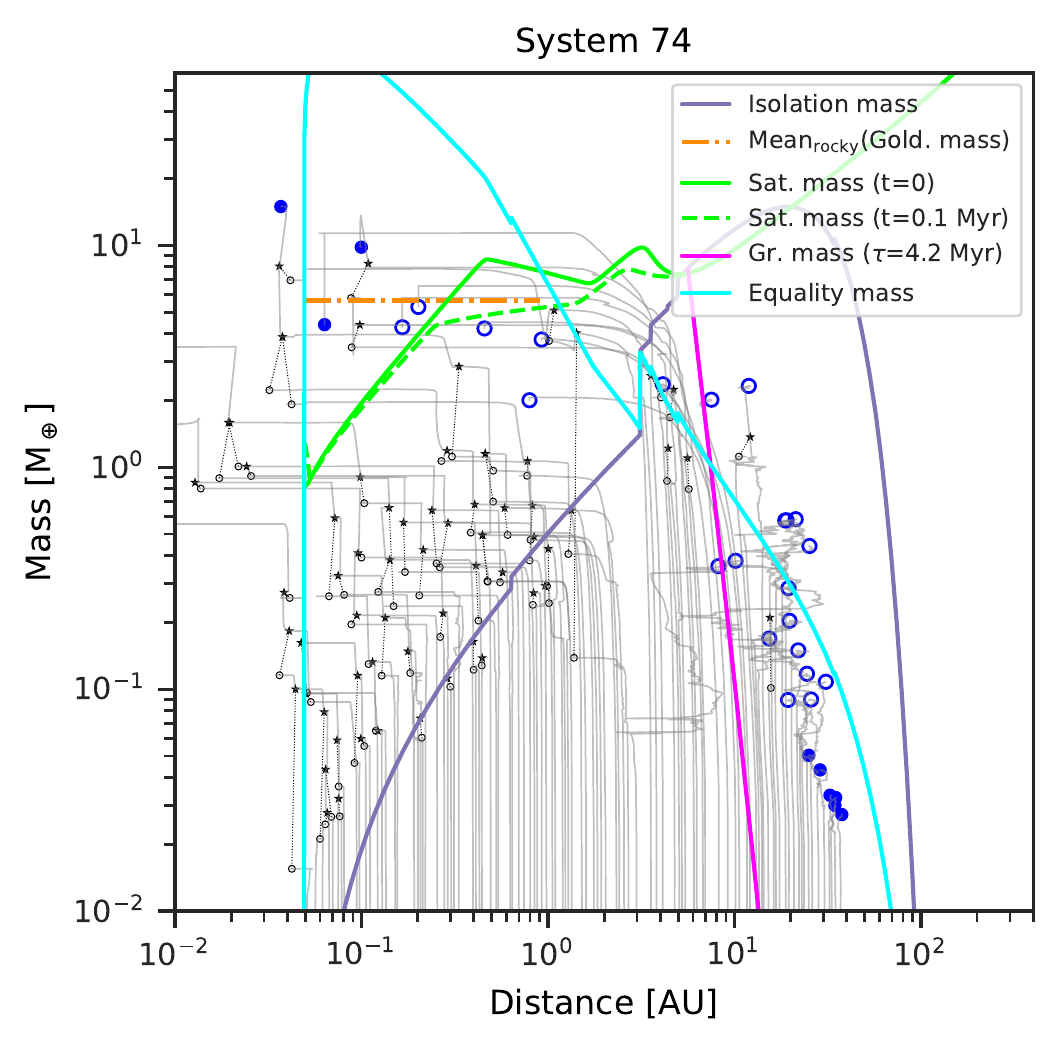}
	\caption{Example of formation tracks and final architecture of a system that belong to Class~II. In this class, the governing mass scale in the inner system is linked to orbital migration and is thus given by the saturation mass $M_{\rm sat}$ where planets are released from migration traps (green lines, evaluated at $t=0$ and $t=\SI{0.1}{\mega\year}$) and/or the equality mass $M_{\rm{eq}}$ where the timescales of oligarchic growth and type I orbital migration  are equal (cyan line).}
	\label{fig:class2}
\end{figure}

\begin{figure}
	\centering
	\includegraphics[width=\textwidth]{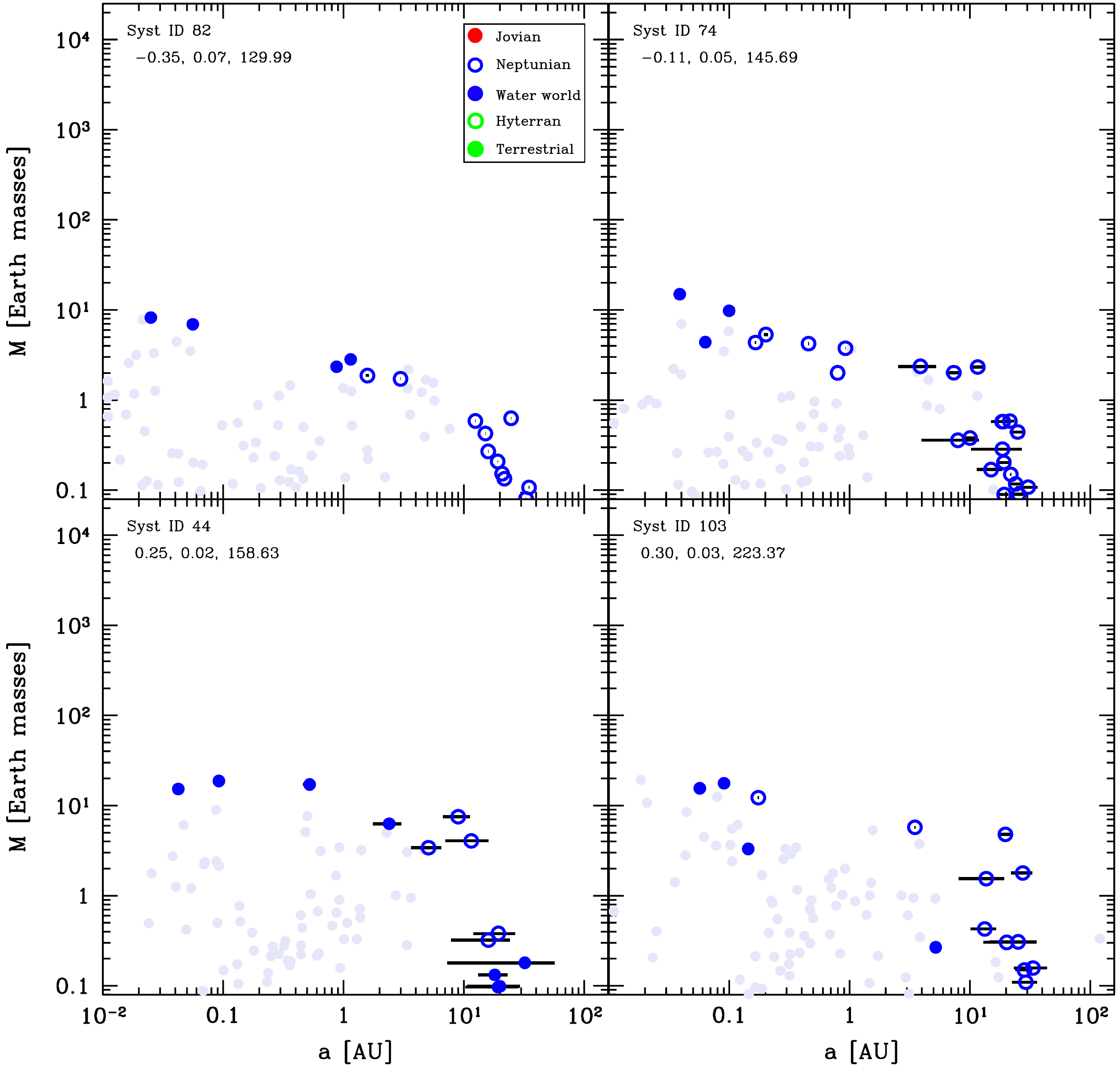}
	\caption{Final architecture of four systems of Class II. The plot is analogous to Fig. \ref{fig:class1examples}. In contrast to Class I, all planets have a volatile-rich composition, and the mass is a decreasing or constant function of orbital distance. One also notes the absence of planets with H/He at small orbital distances. The numerous low-mass planets outside of 10 AU might not have yet settled into their final state after 100 Myr which is the N-body integration time covered in our simulations.}
	\label{fig:class2examples}
\end{figure}

Class II are systems with more massive planets, yet no giants. There are usually 5 to 10 planets whose mass is larger than \SI{1}{\mearth}. In contrast with Class I, most planets are icy, even in the inner region. Another contrast is that the planet masses are either independent of, or anticorrelated with their orbital distance, opposite to Class I.

We show one example of the formation tracks of a system in this class in Fig.~\ref{fig:class2}. It reveals that the inner system is populated by icy planets that originally formed beyond the iceline and then migrate inward at constant mass following similar tracks as the `horizontal branch' planets already identified in \citet{2009A&AMordasinia}. The terrestrial planets that also originally from in the system are captured in mean-motion resonances, migrate in resonant convoys, and end up either pushed into the star or get destabilised, which leads to a collision with an icy planets; they usually do not survive. In our scheme, the collision between a terrestrial and an icy planet result in an icy planet, as we use a threshold of \SI{1}{\percent} of volatiles by mass in the core to classify planets as icy and planets forming beyond the iceline have a volatile mass fraction up to \SI{59.5}{\percent}, which is radially variable and depends on whether the individual volatile species are in condensed or gaseous form. One way to distinguish planets that had a giant impact during the migration phase is to see if they have a H/He envelope. Giant impacts with the terrestrial planets will release a sufficient amount of energy to strip the envelope from the icy planets \citep[e.g.][]{2020MNRASDenman}. This is the case for instance of the first and third innermost planets. The loss of the envelope of the second innermost planet is due to a runaway bloating process leading to Roche lobe overflow. In this process, the radius increases because the heating due to bloating, which further increases bloating. This occurs for close-in low-mass planets with a high envelope mass fraction (like for runaway Roche lobe overflow due to tidal heating, \citealp{2015ApJValsecchi}, or atmospheric escape, \citealp{2004AABaraffe,2022ApJLThorngren}). We have verified that the envelope is also lost by XUV-driven atmospheric photoevaporation in the absence of Roche lobe overflow.

Unlike systems in Class I, there is sufficient mass beyond the iceline to form sub-Neptunes within the lifetime of the protoplanetary disc. The planets forming just outside the iceline are thus able to reach the equality mass (the cyan line in Fig.~\ref{fig:class2}) early enough for the gas disc to be massive enough to exert significant torques and thus migrate inward. We note that some planets experience strong inward migration even below the saturation mass (green lines); this is because the saturation mass decreases over time along with the gas surface density. As consequence, the saturation mass at the time where planet begin to rapidly migrate is lower than the value shown in the figure. Further, the planets do not experience outward migration below the saturation because they are not in a region where the corotation torque is large enough to overcome the Lindblad torques. The planet on the horizontal branch have masses that are too low for rapid gas accretion, meaning that they do not reach the critical (core) mass. The formation of Class~II systems relies on migration. Thus, the physical process responsible for the radial transport of gas is of special importance: in the case of magnetically-driven disc winds combined with low viscosity, migration would likely be reduced \citep{2016ApJSuzuki}, and part of the Class~II systems could behave more like Class~I systems, but with more massive planets. However, the corotation torque depends on the viscosity \citep{2001ApJMasset} and the transport mechanism \citep{2020AAKimmig}, so this would have to be verified by hydrodynamical simulation.

We further provide the final architectures of four more systems in this class in Fig.~\ref{fig:class2examples}. These systems show a common pattern of planets of masses between about \num{2} and \SI{10}{\mearth} planets from the inside to about \SI{2}{\au}. This is a relic of the initial growth beyond the iceline. The equality mass (cyan line in Fig.~\ref{fig:class2}) is a strongly decreasing function of distance, which is mainly linked to the increase of the planetesimal accretion timescale with distance. Planets just outside of the iceline will thus be able to reach larger masses before being subject to significant migration. Further, as the the growth timescale increases with distance, the order of the planets is retained through the migration stage; with the earlier and more massive planets undergoing stronger migration than the more distant ones. Therefore, the more massive planets origination from just outside of the iceline migrate in first, followed later on by less massive planets originating from further away. The resulting effect of mass decrease with distance is further compounded by the growth by giant impact during the migration. As we pointed out above, the innermost planets are H/He-free while more distant ones have some gaseous envelope, and this is the consequence of giant impacts stripping the envelope of the inner planets. The envelope represent only a fraction of the total mass, so the total mass change by giant impact is usually dominated by the gain of core mass rather than the loss of the envelope. Planet masses is not a strict discriminant between Class~I and Class~II systems. For instance, the innermost planets in the system shown on the top left panel of Fig.~\ref{fig:class2examples} are about \SI{10}{\mearth}, while that in the bottom right panel of Fig.~\ref{fig:class1examples} are slightly more massive. The difference lies in that the planets in the Class~II system migrate further than the ones in Class~I, and this is related to the disc lifetimes of the respective systems. We shall discuss this effect in greater details in Sect.~\ref{sec:res-env}.

\subsubsection{Class III: mixed systems with low-mass and giant planets}

\begin{figure}
	\centering
	\includegraphics[width=\textwidth,height=17cm]{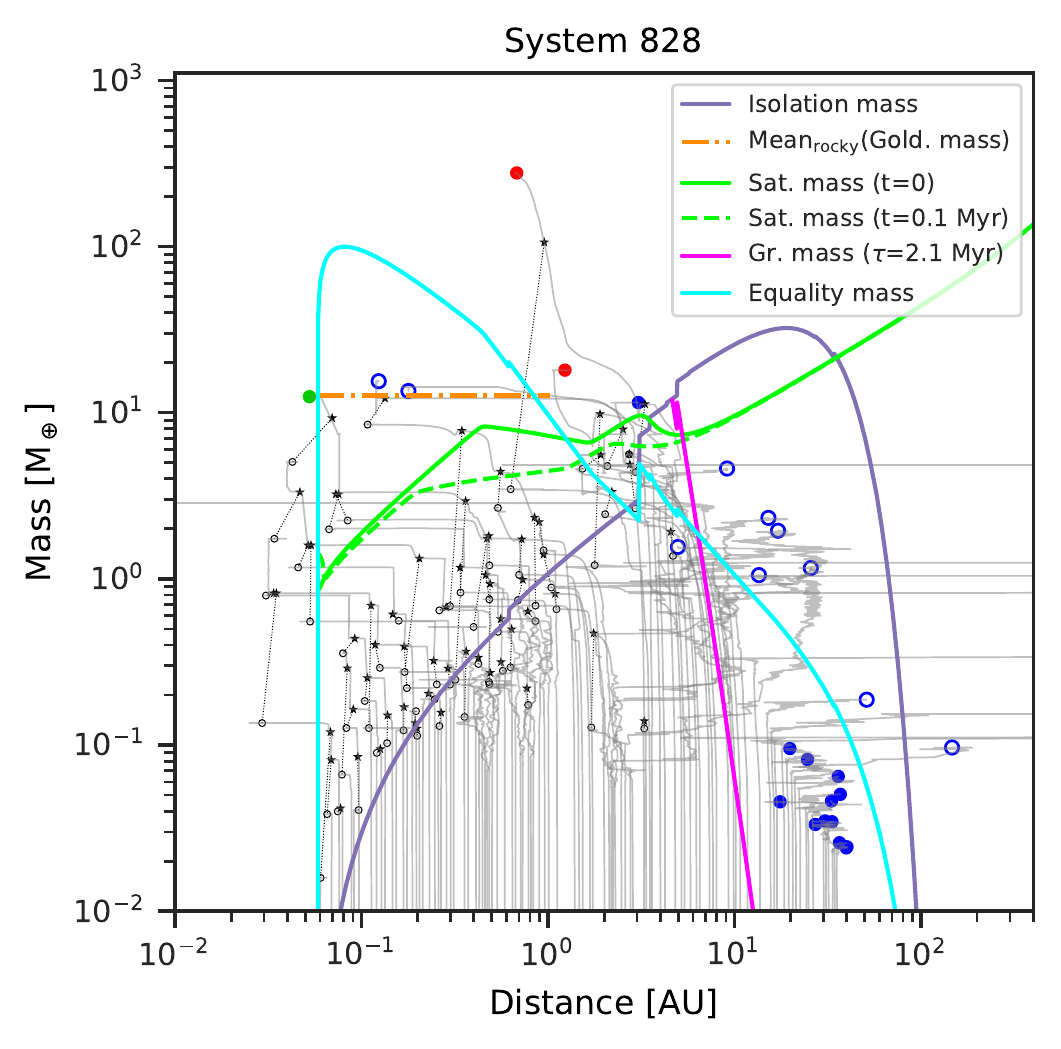}
	\caption{Example of formation tracks and final architecture for a system that belongs to Class~III. In this class and in contrast to Class II, a protoplanet (with typical origin outside the water iceline) manages to grow massive enough to start runaway gas accretion, i.e., it reaches the critical core mass. It then passes into slower Type II migration and grows to a giant planet mass. The final mass of the giant planet is in this system similar to the one of Jupiter, but is found at about 0.7 AU. The system also contains several lower mass planets in- and outside of the giant planet. During its gas accretion phase, the giant accreted a 3 $M_\oplus$ planet. 8 protoplanets were ejected out of the system and one fell into the star, as visible by grey lines crossing the right and left y-axes, respectively.}
	\label{fig:class3}
\end{figure}

\begin{figure}
	\centering
	\includegraphics[width=\textwidth]{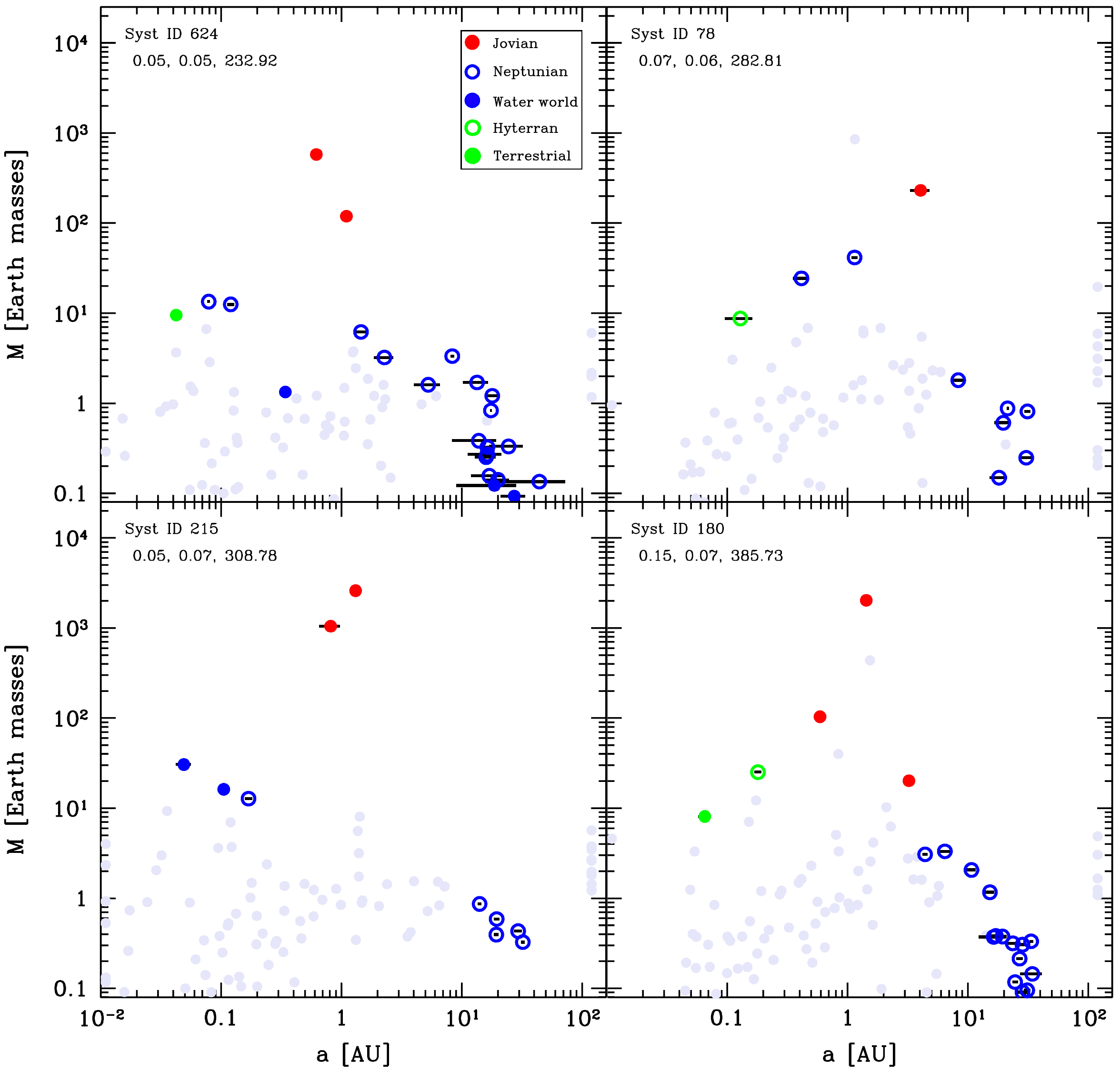}
	\caption{Final architectures of four systems of Class III, analogous to Fig. \ref{fig:class1examples}. The diversity is very large regarding both the orbital configuration, mass, and number of the giant planets but also regarding the composition of the inner low-mass planets which can be of a Neptunian, water world, hyterran or terrestrial type. }
	\label{fig:class3examples}
\end{figure}

Class III systems have giant planets, along with inner less massive planets. They also have outer lower mass planets. Here again, we provide formation tracks of one system in this class in Fig.~\ref{fig:class3}. The initial formation stage is similar to system in Class~II, with the embryos inside the iceline accreting in-situ up to the isolation mass, while the embryos beyond the iceline begin to migrate before reaching the isolation mass. The contrast with systems in Class II is that the planets beyond the iceline continue to grow by gas accretion during their inward migration and end up undergoing runaway gas accretion.  At this point, the track of the forming giant planet bends strongly upwards. The reason is the higher (equality) mass these proto-giants have (close to \SI{10}{\mearth}, instead of a few Earth masses), as discussed below. Once the planet undergo runaway gas accretion, the migration rate decreases in part because in the inner disc, the migration rate is limited by the ratio between the planet mass and that of the local gas disc. As a consequence, the giant planets migrate less than the sub-Neptunes in Class II, and leave the inner part of the system free from perturbation. The inner system behaves with features of both Class~I and~II. The part resembling to Class~I is that there remain terrestrial planets that grow by giant impacts, while the part resembling to Class~II is the presence of two Neptunian planets inside of the giant planets that underwent migration.

The initial disc of the system presented in Fig.~\ref{fig:class3} is more massive than that in Fig.~\ref{fig:class2}. As a consequence, the isolation mass is also larger while the growth timescale $\tau_\mathrm{grow}$ is smaller, meaning faster accretion. This pushes the equality mass to larger values, allowing the planets to reach larger masses for a given amount of migration compared to the Class~II system. The larger core masses allow the embryos to bind proportionally even more gas (as the Kelvin-Helmholtz cooling timescales decreases for larger body masses, \citealp{2000ApJIkoma}), allowing them to undergo runaway gas accretion on a time that is less than that of the inward migration.

Four final systems in this class are shown in Fig.~\ref{fig:class3examples}. As they show, the inner planets can be any combination of rocky or icy bodies. In systems with both rocky and icy inner planets, the same compositional ordering as in Class~I is retained. Rocky planets have a formation pattern similar to those in Class~I, while the inner icy planets have similar formation patterns as for those in Class~II, migrating after having accreted from outside the iceline. This also means that the giant planets are not necessarily the first embryos just beyond the iceline. Outside of the giants, we see groups of several planets with masses of a few \si{\mearth}. They consist of mostly ices and some H/He, reminiscent of the Solar System ice giants. They still grow slowly from remaining planetesimals. Over long timescales, they might coalesce to form more massive ice giants.

One common feature between the different systems in this category is that they are dynamically cool. Planet eccentricities remain relatively low, even when multiple giant planets are present. Still, there are low and mid-mass planets in the outer regions that get on wide orbits or ejected (but not the giant planets themselves). However, this is limited to protoplanets outside the position of the giants.

\subsubsection{Class IV: Dynamically-active giants}

\begin{figure}
	\centering
	\includegraphics[width=\textwidth,height=17cm]{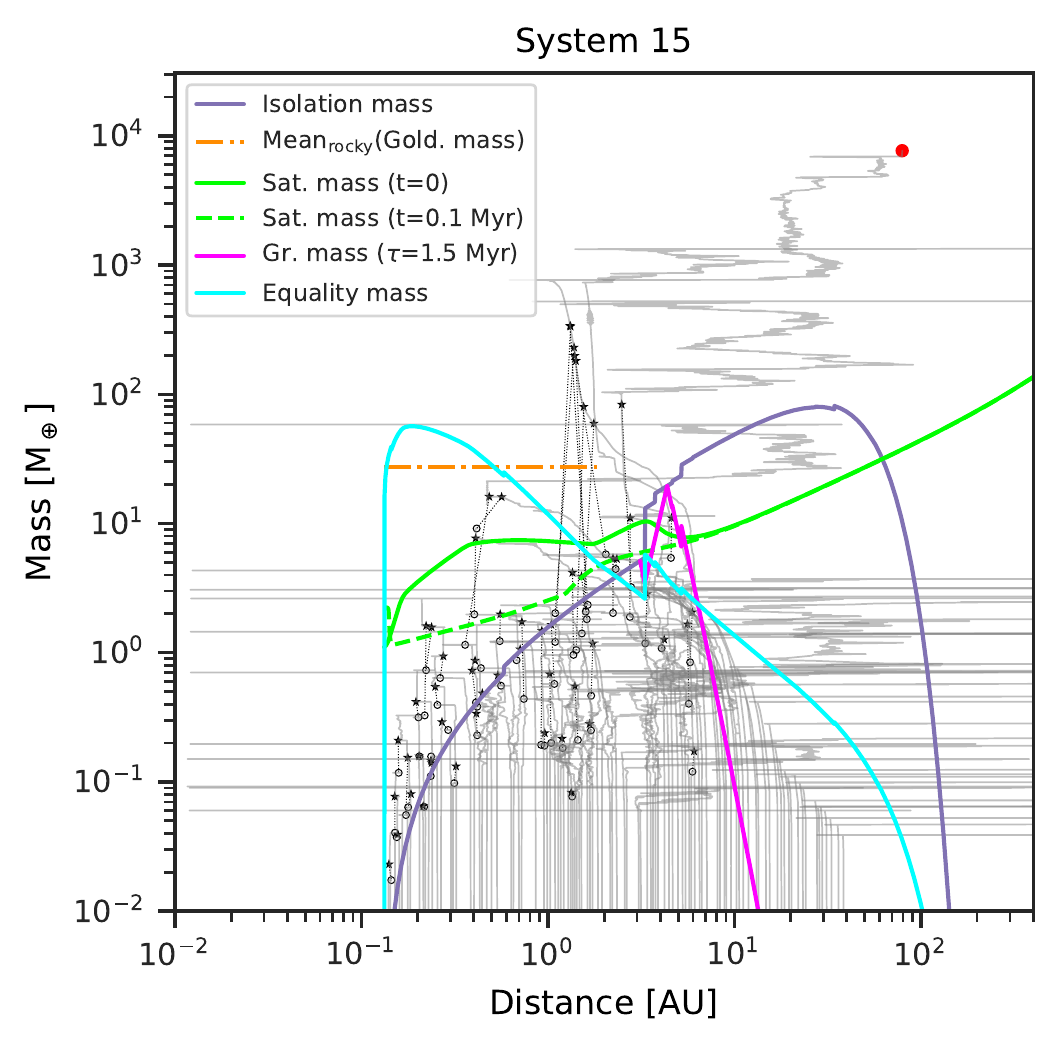}
	\caption{Formation tracks of a system that belongs to Class~IV. The concurrent formation of several (proto)giant planets (bodies reaching the critical core mass for runaway gas accretion) in spatial proximity lead to violent dynamical interactions among these massive planets, strongly disturbing the lower-mass bodies in the system via collisions and ejections. In the end, in this example only one very massive, distant and eccentric giant planet remains. }
	\label{fig:class4}
\end{figure}

\begin{figure}
	\centering
	\includegraphics[width=\textwidth]{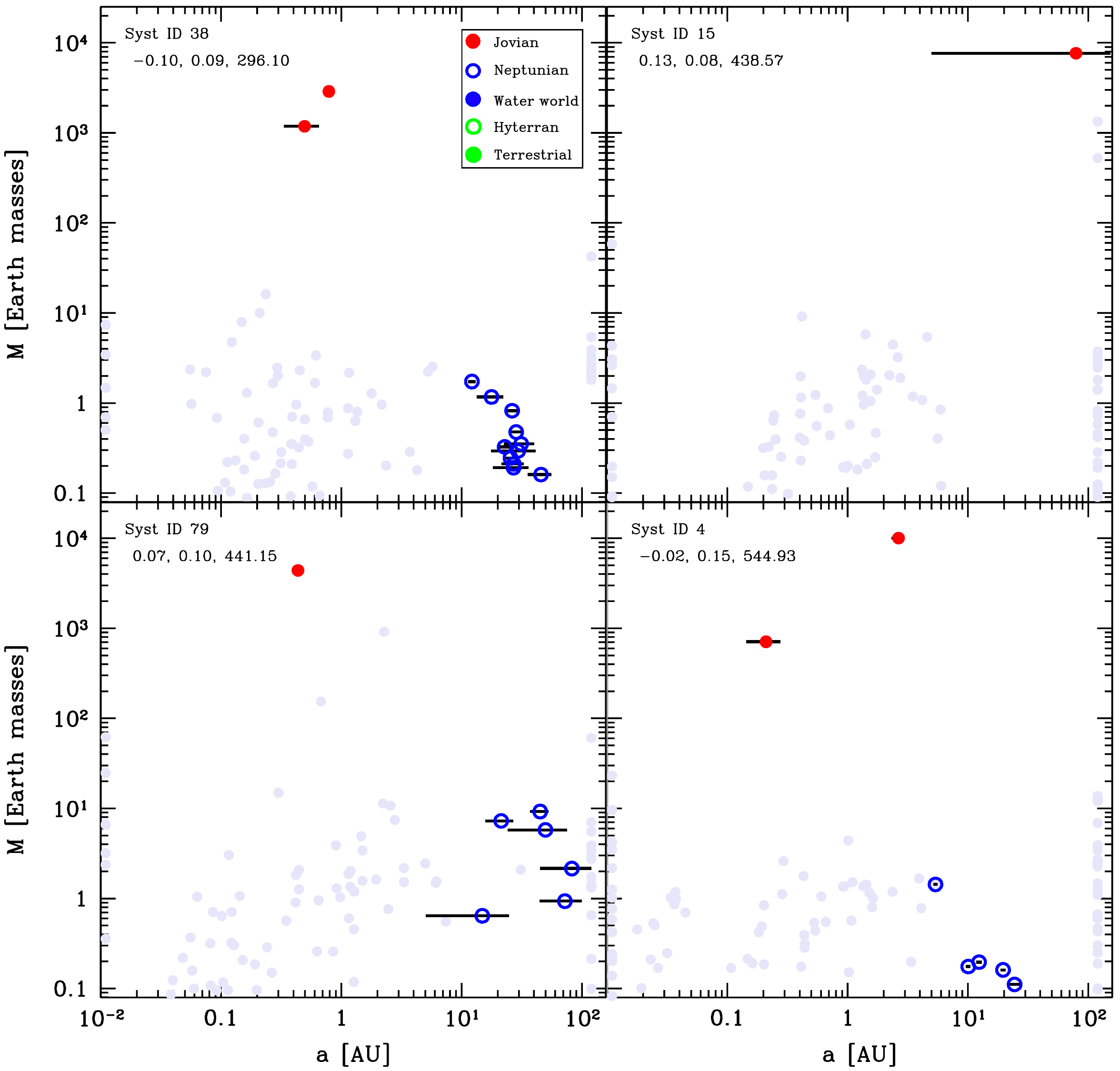}
	\caption{Final architectures of four systems of Class IV, analogous to Fig. \ref{fig:class1examples}. From the points next to the left and right y-axis, one notes the significant number of planets that have either been ejected out of these systems or that have collided with the host star. In the system in the bottom left, collisions among giant planets have also occurred. }
	\label{fig:class4examples}
\end{figure}

Systems in Class IV also have giants, like Class III; the difference lies in that they do not contain inner low-mass planets. Overall, they have a small number of remaining objects (sometimes even only one or two), with possible low-mass bodies at large distance. Systems in this class tend to have very massive giant planets (\SI{\sim10}{\mj}), although this is not always the case.

As for the other classes, we provide one example of the formation history of a system in this class in Fig.~\ref{fig:class4}. It reveals that the system formed a total of three giant planets, two of which have subsequently been ejected, leaving a single giant on a wide and eccentric orbit. These planets begin their formation similar to the system in Class~III discussed above, but the giants dynamically interact and get destabilised. The instability arises if several protoplanets in close-proximity start gas runaway accretion within a short time interval. It should be noted that our model assumption of inserting all embryos at one moment (at t=0) may artificially lead to such situations. If the early phases of solid growth (from dust via pebbles to planetesimal and embryos) are included, bodies may emerge more gradually \citep{2022AAVoelkel}. The dynamical instability caused by the giant planets affects the other bodies in the system, starting with close-by bodies, some of which get accreted by the innermost giant. The others are sent on very eccentric orbits, ending up either colliding with the star or ejected.

We also provide the final state of four systems in this class in Fig.~\ref{fig:class4examples}. While they show a pattern of one or more giant planets, these planets have a diversity of configurations, from compact pairs inside \SI{1}{\au} (top left panel) to a single body at \SI{\sim100}{\au} (top right panel, the same one as in Fig.~\ref{fig:class4}). In addition, other low-mass objects at large distances may remain. These are the remainder of the outer embryos which have not been affected by the instabilities of the inner giant planets. For instance, none remain in system \texttt{15} (top right panel) because the planet was sent to an eccentric orbit with large separation while two other giants have been ejected (as revealed by the grey dots at \SI{100}{\au}); however, in the other systems there are no giant planets either ejected or on wide orbits, leaving the outer region free from perturbation that would destabilise these bodies.

One of the main difference between systems of Class III and IV are the dynamical instabilities. The presence of very massive giants does favour the occurrence of planet-planet scatterings, and thus systems with the most massive giants are more likely to be in Class IV.

\subsection{Key mass scales in planetary systems}

\begin{figure}
	\centering
	\includegraphics[width=0.49\textwidth]{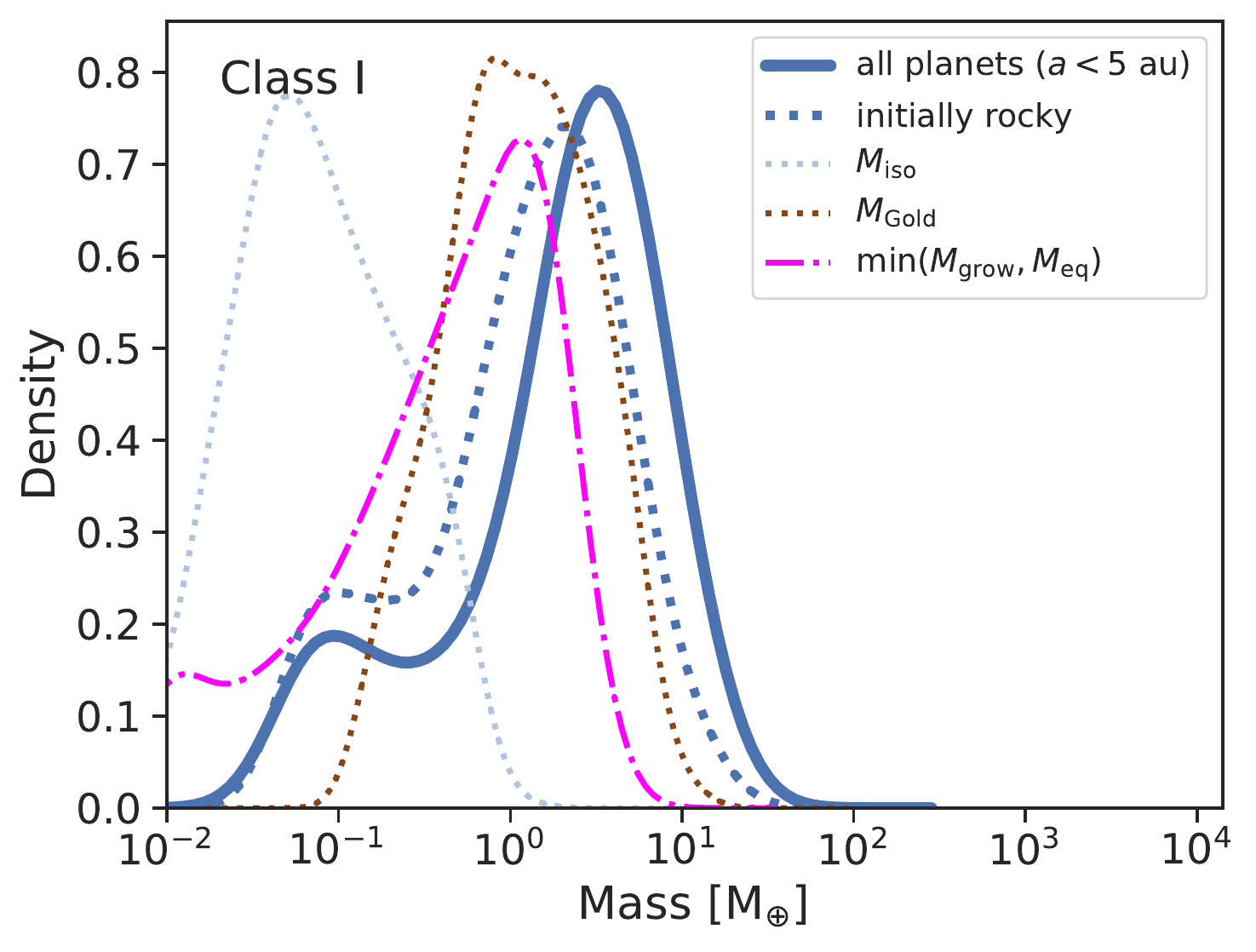}
	\includegraphics[width=0.49\textwidth]{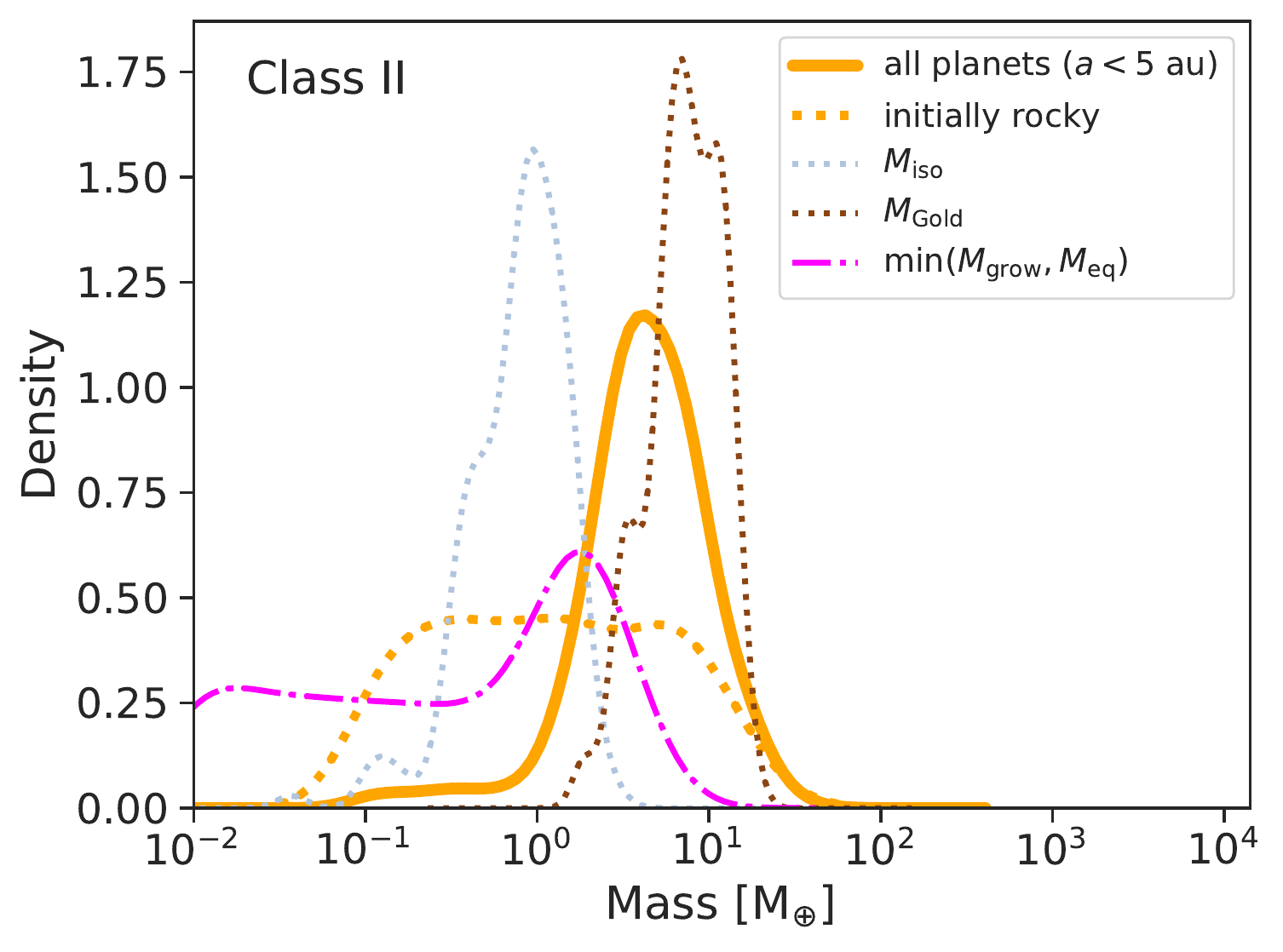}
	\includegraphics[width=0.49\textwidth]{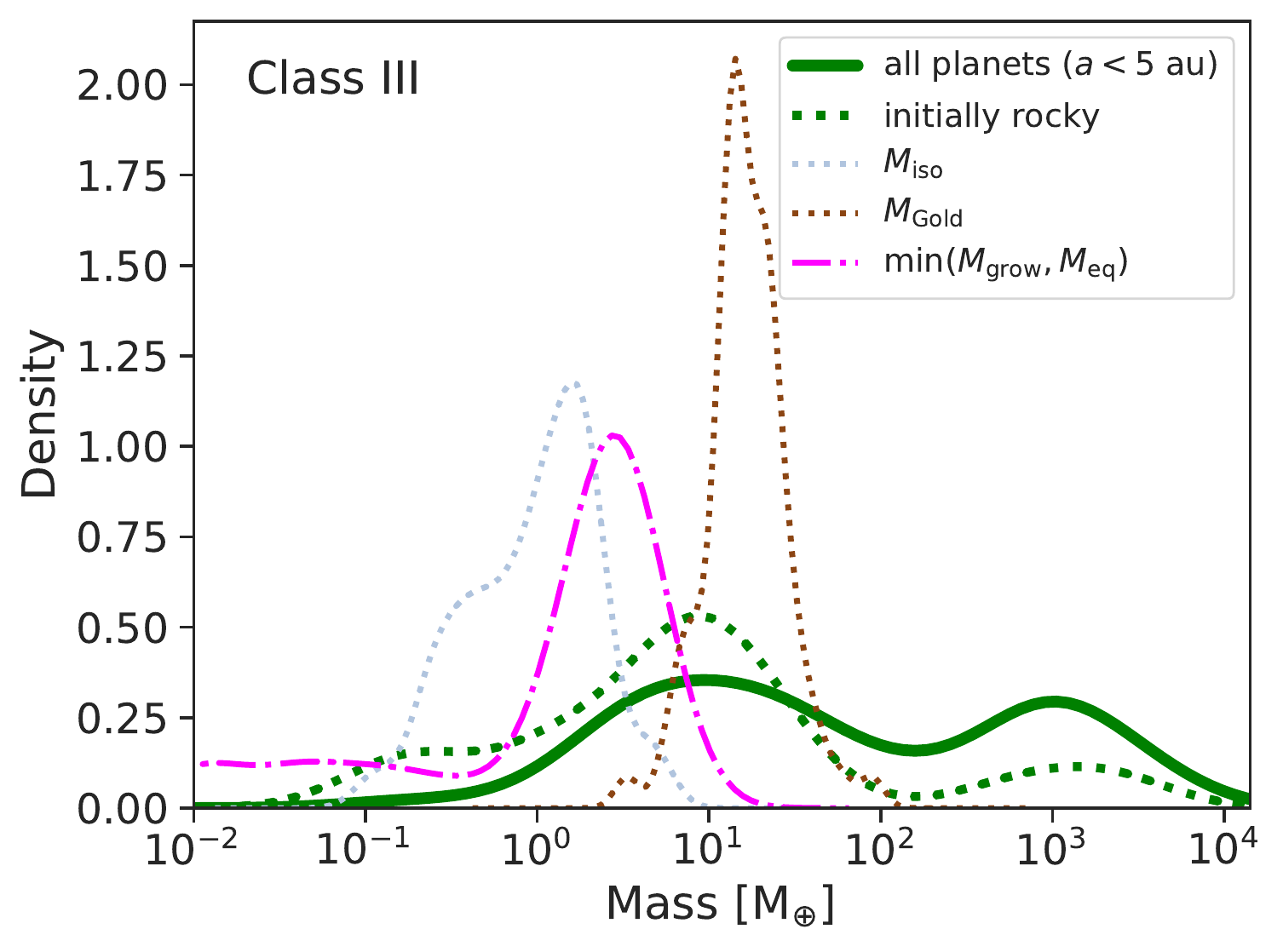}
	\includegraphics[width=0.49\textwidth]{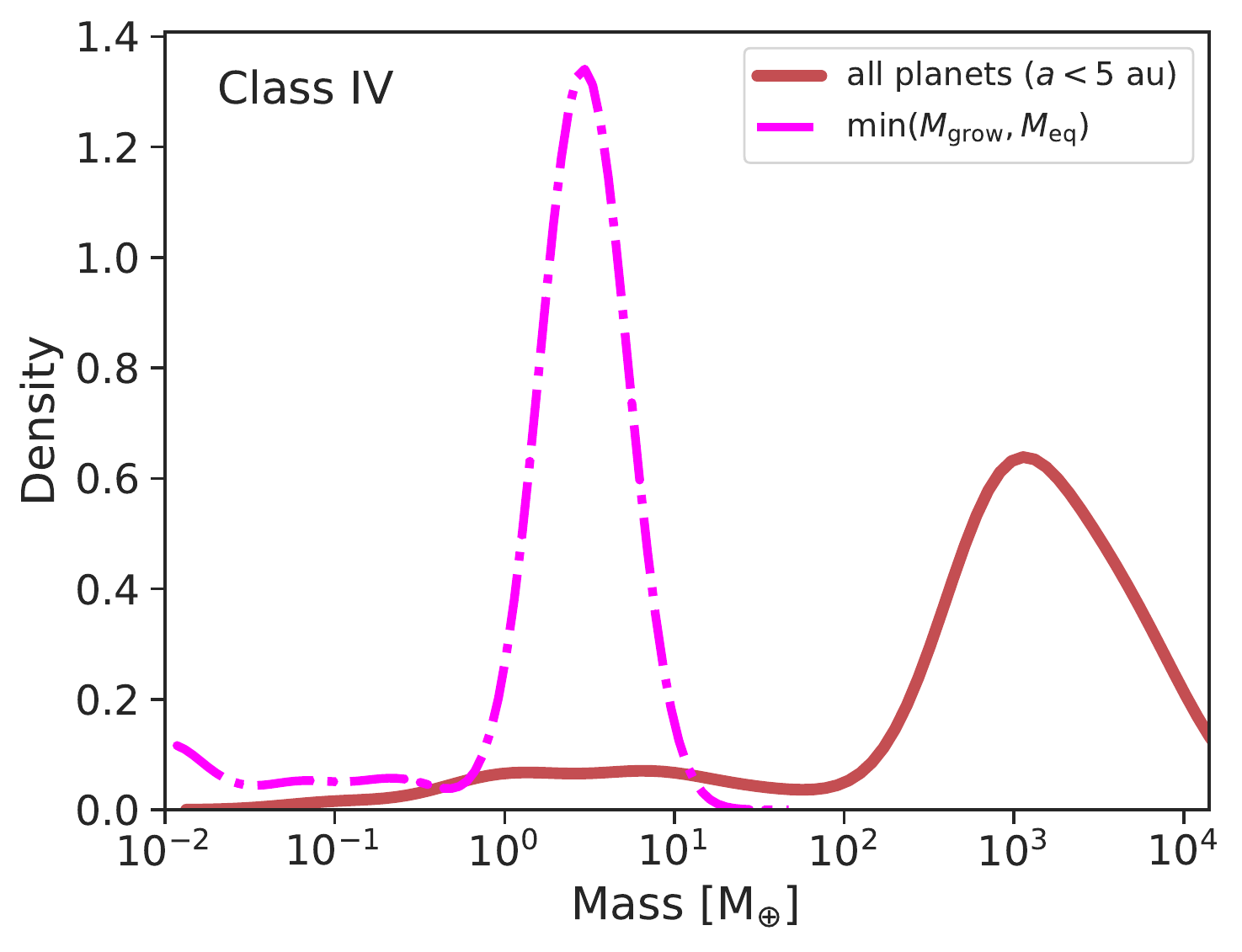}
	\caption{Gaussian kernel density estimates of the planet mass function for planets within \SI{5}{au}. For each line, the same bandwidth of 0.3 dex was chosen. The isolation (Eq.~(\ref{eq:miso})) and Goldreich (Eq.~(\ref{eq:mgold})) mass scales are plotted only for initially rocky planets (dotted lines). Most of the planets in the sample were however initially icy. For comparison to those, we show with the magenta curve the smaller of the growth $M_{\rm grow}$ (Eq.~(\ref{eq:mgrow})) or the equilibrium mass $M_{\rm eq}$ (Eq.~(\ref{eq:meq})). Due to the lack of statistics (only five planets), we omitted the rocky planets and their mass estimates for Class IV.}
	\label{fig:mass_func_classes}
\end{figure}

In Sect.~\ref{sec:timescales} we introduced a number of analytical key mass scales which can be used to understand the main mechanisms of planetary formation. We then used them to understand the formation tracks of individual systems. Here, we expand this analysis to the population as a whole. Figure~\ref{fig:mass_func_classes} shows the comparison of the various mass scales with all the planets in the different classes of planetary systems.

It is interesting to see that the typical mass scales can predict to an order of magnitude or better the modelled distribution of solid-dominated masses. For the icy planets, we expect them to be limited by either running out of time to grow or by starting to migrate. The smaller of the growth or equality mass is shown in Fig.~\ref{fig:mass_func_classes} as the magenta line. This simple argument predicts a bimodal distribution of `failed' cores at the initial mass of \SI{0.01}{\mearth} and at a larger mass varying from \SIrange{1}{10}{\mearth}. The actual upper peak shown as the dash-dotted lines are for Class I and II located a factor of \num{\sim4} above this simple estimate. From evolution tracks it becomes evident that planets in this category can be trapped at the edges of outward migration regions or undergo giant impacts which enhances their mass.

The fact that collisions between embryos take place is factored into the Goldreich mass, which we however only applied to the inner rocky planetary system, where the simulation time is long compared to the growth timescale. The rocky planets in Class I systems are also bimodally distributed. There, we see that the lower peak is located close to the isolation masses while the upper is well reproduced by the Goldreich mass. This can be interpreted as systems or planets which went into a dynamically active stage leading to a giant impact phase forming the upper peak while lower mass planets are also predicted by the simulations remaining at their location without excitation and the possibility to grow from solid material outside their feeding zones.

For Class II systems with significant migration patterns, the low-mass rocky planets remain slightly below the mean isolation masses while the upper peak is reduced. Also, the overall number of initially rocky planets existing at the end of the simulations is reduced, a trend which continues also for the classes with giant planets.

This reduces the statistical significance and meaningfulness of the comparison for the rocky planets. Furthermore, for Classes III and IV, the final planetary systems are dominated by the massive, gas-rich planets which leads to expected deviations from the simple, solid accretion based mass scales. For gas-dominated planets, see \citet{2022ApJAdamsBatygin}.

\subsection{Links between initial conditions, system properties, and architecture classes}
\label{sec:res-linkiniconds}

Now, we aim to understand the conditions needed for the formation of the different kinds of planetary systems. We find that among the different initial conditions of our simulations, the initial mass in solids $M_{\rm p,ini}$ is the one that can be used best to discriminate the different classes of systems. However, it is not the only discriminant.

\begin{figure}
	\centering
	\includegraphics[width=0.99\textwidth]{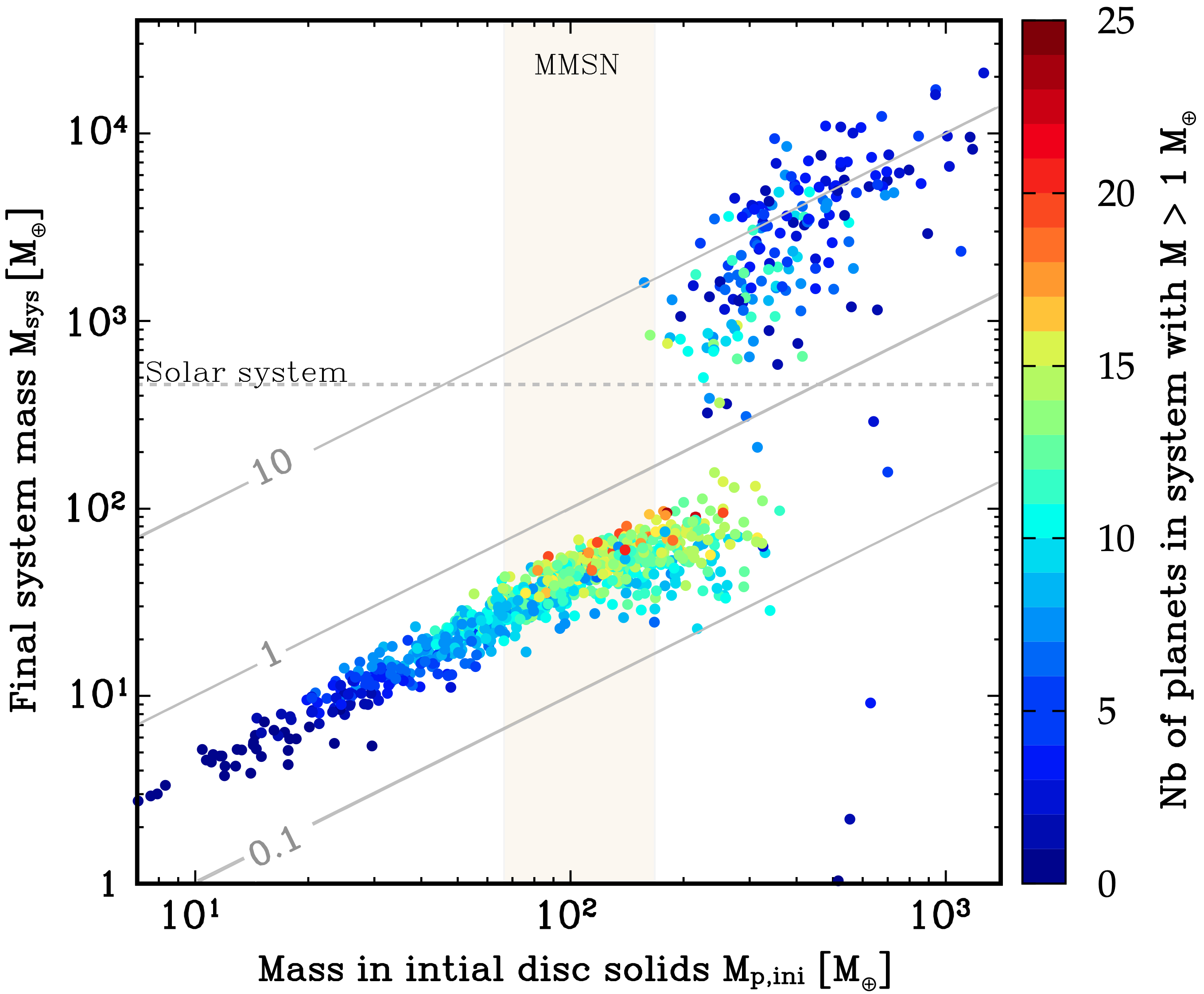}
	\caption{Final mass of the planetary system (i.e., sum of all planets in a system at \SI{5}{\giga\year}) $M_{\rm sys}$ as a function of the initial mass of the solids (planetesimals) in the parent disc $M_{\rm p,ini}$. The colours show the number of planets in each system with a mass of at least \SI{1}{\mearth}. From this colour code, we see how the number of planets as a proxy of system architecture systematically changes with the initial mass in solids. The vertical shaded region `MMSN' shows the estimated initial minimum solid content of the Solar Nebula. The horizontal dotted line `Solar System' is the total system mass today (\SI{\approx450}{\mearth} = essentially the sum of the masses of the four giant planets). Diagonal lines show conversion efficiencies (which can be larger than unity because accreted H/He is included in the system masses, but not in $M_{\rm p,ini}$).}
	\label{fig:mass_disc_system}
\end{figure}

One way of illustrating the impact of the initial mass in solids is shown in Fig.~\ref{fig:mass_disc_system}. It shows for all 1000 synthetic planetary system as a function of the initial mass of solids (planetesimals) in the natal disc $M_{\rm p,ini}$ the final mass of the planetary system $M_{\rm sys}$ (i.e., sum of all planets in a system at \SI{5}{\giga\year}).

For this system mass, one can clearly distinguish the systems without ($M_{\rm sys}\lesssim \SI{150}{\mearth}$) and with one or several giant planets ($M_{\rm sys}\gtrsim \SI{150}{\mearth}$). As expected for a formation model based on the core accretion paradigm \citep[e.g.][]{2012A&AMordasiniA}, giant planets only form in discs with sufficiently high $M_{\rm p,ini}$ of about 150 to \SI{200}{\mearth}. For the systems without giants, there is a clear linear scaling between $M_{\rm p,ini}$ and $M_{\rm sys}$ with an efficiency factor of about \SI{30}{\percent}. For the systems with giants, there is still a correlation, but it is less clear. This is expected, because for the final masses with giant planets the disc gas mass is important and not the solid mass. There are also some special cases  with very small $M_{\rm sys}$ for high $M_{\rm p,ini}$. They are discussed below.

The colours in Fig.~\ref{fig:mass_disc_system} show the number of planets with a mass of at least \SI{1}{\mearth} in each system (including planets at all semi-major axes). This planetary multiplicity is a proxy for system architecture and exhibits an interesting pattern. At the lowest $M_{\rm p,ini}$ of about \SI{10}{\mearth}, not even a single planet with a mass of at least \SI{1}{\mearth} forms, because the planets remain at sub-Earth masses. As we move to higher $M_{\rm p,ini}$, but still below the value needed for giant planet formation, the number of planets increases. At $M_{\rm p,ini}$ between 100 and \SI{200}{\mearth} there are some planetary systems with a very high multiplicity of up to 25. In this region, we see also a vertical stratification where at given $M_{\rm p,ini}$, systems with a lower $M_{\rm sys}$ also have a lower number of planets. This is a consequence of another disc property, the disc lifetime, which is influenced mostly by the initial mass of the gas disc and the external photoevaporation rate. At given $M_{\rm p,ini}$, system with a longer disc lifetime have a lower $M_{\rm sys}$. The reason is not that these systems form less planets (smaller mass or lower number) in the first place, but that planets have more time to migrate. This has the consequence that more planets fall into the star both by migration in resonant convoys during the presence of the gas disc, or on long timescales after disc dissipation via tides. For the latter the close-in parking location in systems with significant inward migration makes the difference. This reduces both $M_{\rm sys}$ and the number of planets. It also influences the classification of the architecture, as we will see soon.

Moving finally to the $M_{\rm p,ini}$ that are high enough to allow giant planet formation, we see that the number of planets again decreases. This is a consequence of the fact that growing giant planets can destabilise lower mass planets in their vicinity. These planets are then either ejected from the system or accreted by the growing giant planet(s). This can be seen in Fig.~\ref{fig:class3} where the star symbols on the tracks of the growing giants indicate the accretion of lower mass planets during the process of rapid gas accretion. At $M_{\rm p,ini}$ less than about \SI{400}{\mearth}, there are however still many systems with about 10 planets. Violent instabilities among the giants are particularly efficient in removing lower-mass planets. Such cases can be seen at the highest $M_{\rm p,ini}$. For these systems, the number of planets reaches---as for the lowest $M_{\rm p,ini}$---again very low values of just one or two (but now very massive) planets.

\subsubsection{Solar System analogues}

Figure~\ref{fig:mass_disc_system} also shows the location of the minimum-mass Solar nebula (MMSN; with the boundaries discussed in \citealp{2021AAEmsenhuberB}: $M_{\rm p,ini}$ of 66 to \SI{151}{\mearth}) and the final Solar System. The Solar System is not straightforward to form in our current model. This is mainly because gas accretion is assumed to remain unaffected by gap formation. Planets undergoing runaway gas accretion therefore will rapidly attain a Jupiter mass or more, leading to systems that are more massive than the Solar System. Obtaining more giant planets whose mass is compatible with that of Jupiter or Saturn would require lower mass accretion rates \citep{2018ApJSuzuki,2019MNRASNayakshin}. Low disc viscosities \citep[][and references therein]{2018ApJSuzuki} coupled with the inclusion of gap formation could alleviate this problem. Further, the model does not produce many systems with giant planets for discs whose mass are in the MMSN range. Rather, $M_{\rm p,ini}$ that are a factor \num{\sim2} more massive are need. This is linked to the overall conversion efficiency of the model, where only about a third of the solids are accreted onto the planets, as it can be seen from the many systems without giant planets. In the outer part of the disc, accretion timescale is larger than the disc lifetime (see the pink curve in Figs.~\ref{fig:class1}, \ref{fig:class2}, and~\ref{fig:class3}) so only a fraction of the mass is converted into planets. On the other hand, the MMSN is a minimum mass indeed, so the actual $M_{\rm p,ini}$ of the Solar System might had been higher.

In addition, we find that that planets in Solar-system analogues are too close-in (Fig.~\ref{fig:class3examples}). This is in part due to the difficultly of forming distant planets with planetesimal accretion \citep[e.g.][]{2011ApJRafikov}. There are several others aspects in the model which are missing  for the specific case of the Solar System. One is that the model does not include the \citet{2001MNRASMassetSnellgrove} mechanism that leads to outward migration of a pair of giant planets trapped in a mean-motion resonance and sharing a common gap, which has been invoked for the formation of Jupiter and Saturn \citep[e.g.][]{2011NatureWalsh}. Further, we do not take into account that by blocking or at least reducing  the pebble flux, giant planet formation can affect the formation of planetesimals, which later form the terrestrial planets \citep{2015IcarusMorbidelli}.

\subsubsection{Architecture class}\label{sec:architectureclass}

\begin{figure}
    \centering
    \includegraphics{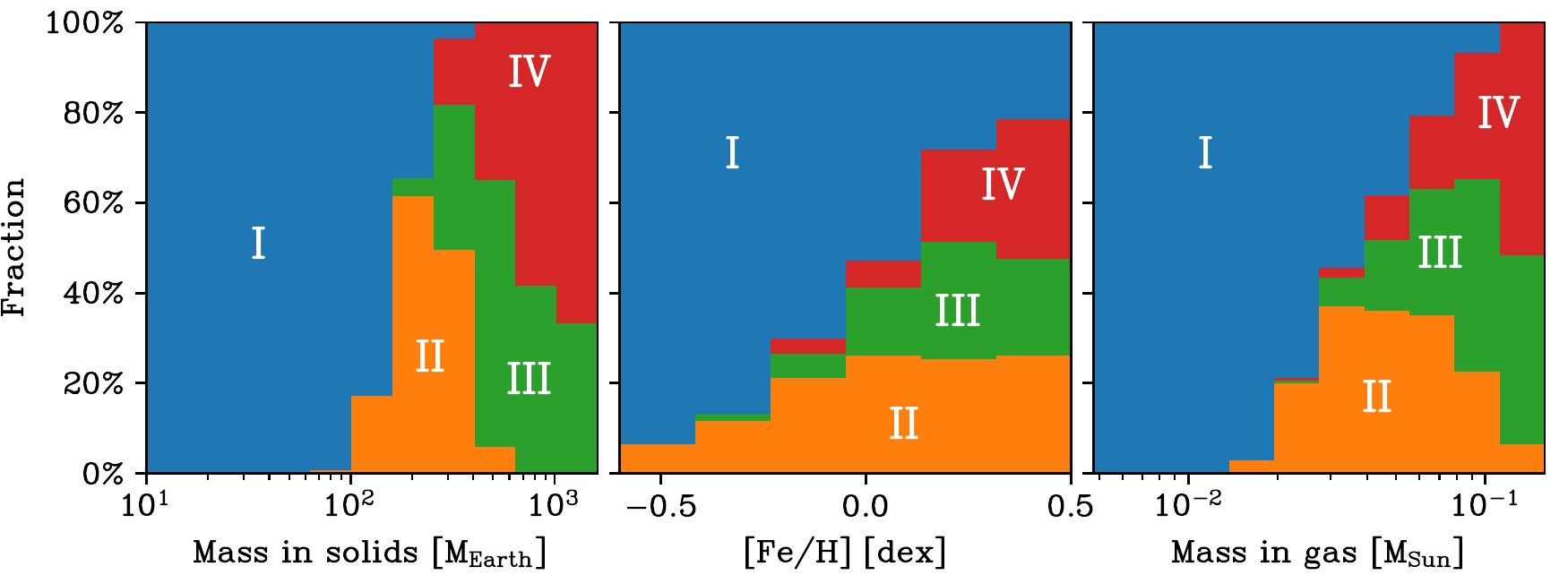}
    \caption{Histograms of the fractions of the different system class as function of initial mass in solids (left), metallicity (centre), and initial mass in gas (right). Blue represents Class~I, orange represents Class~II, green represents Class~III, and red represents Class~IV. The mass in solids is the quantity that most clearly discriminates the classes.}
    \label{fig:sys_class_hist}
\end{figure}

\begin{figure}
    \centering
    \includegraphics[width=0.495\textwidth]{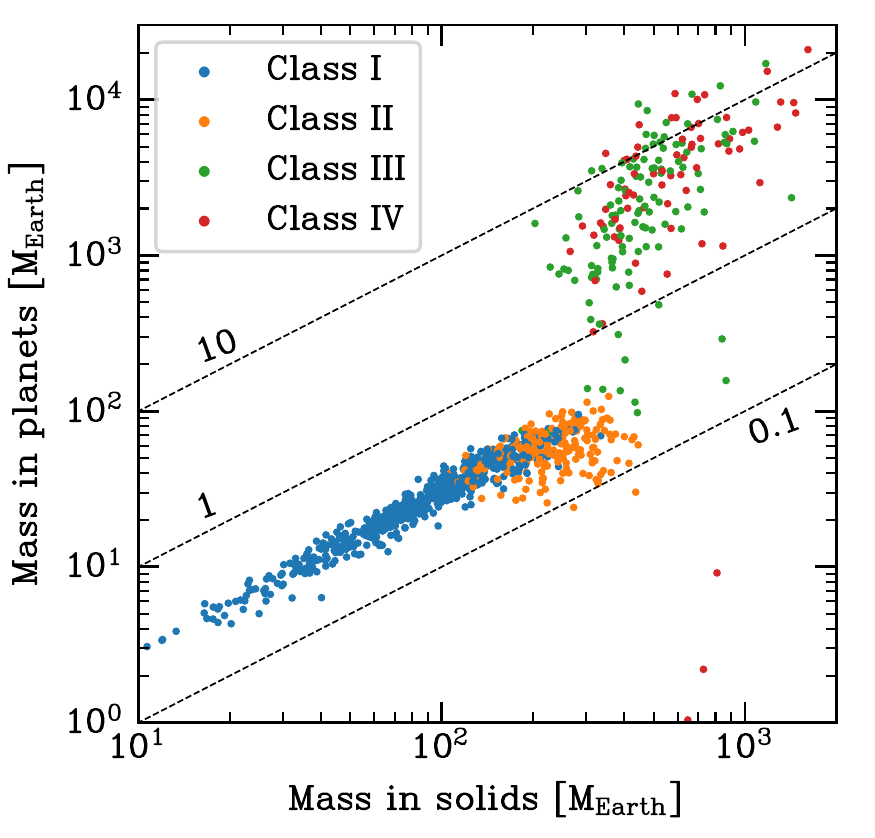}
    \includegraphics[width=0.495\textwidth]{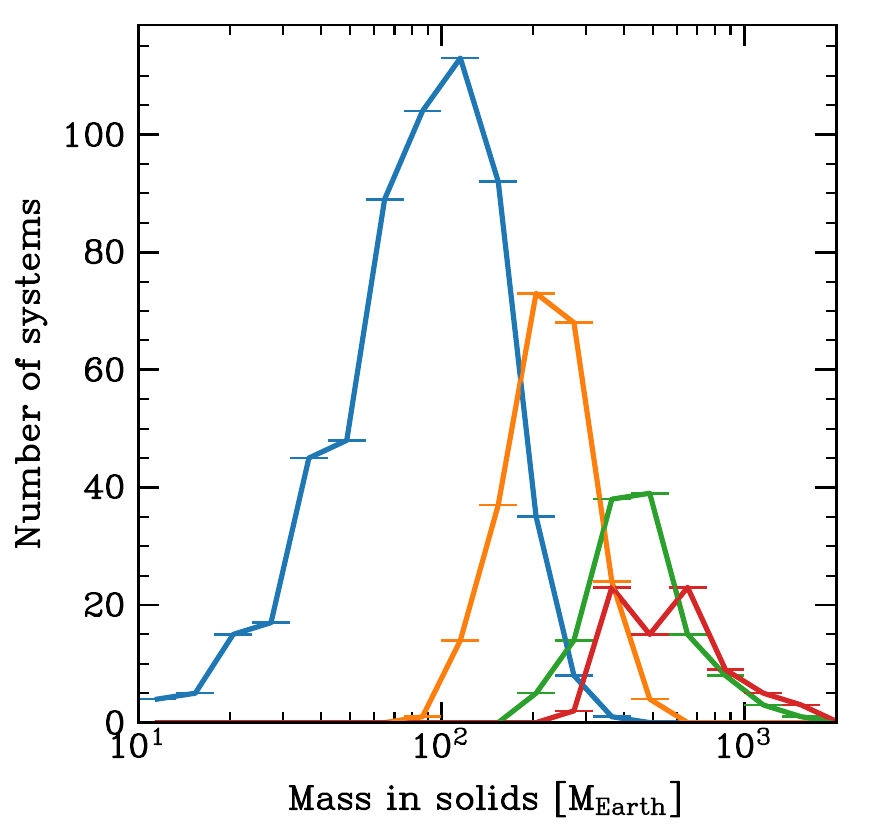}
    \caption{System architecture as function of the initial solids mass content in the disc $M_{\rm p,ini}$ and the total mass (accounting for solids and gas) in the final planets per system $M_{\rm sys}$ (left) and a histogram of initial solids mass only (right). In the left panel, the three dashed diagonal lines again denote the conversion efficiency from disc solids to planets, with the value given next to them. In the right panel, the small horizontal lines show the extent of each bin.}
    \label{fig:sys_dsm_ptm}
\end{figure}

Instead of considering the number of individual planets in a system as a proxy of system architecture, we can also study the four classes directly. We illustrate the effect of different initial disc properties on architecture in Fig.~\ref{fig:sys_class_hist}, with histograms of the fraction of system class as function of solid mass, metallicity, and gas mass. We find the mass in solids is the quantity that best discriminates the different categories, having the different classes best confined to certain regions. The effect of the solid disc mass is reflected on the metallicity and gas mass, since the three quantities are related (only two of the three can be freely chosen). The middle panel shows that Class~I systems are anti-correlated with metallicity, having the largest fraction of systems at low metallicity and steadily decreasing, while the other three quantity show an initial increase following by a plateau. The gas mass follows a similar pattern as the solid mass, but with more overlap between the different classes.

In Fig.~\ref{fig:sys_dsm_ptm} we focus again on the solid mass, by again showing the total mass in planets $M_{\rm sys}$ in each system is  plotted against the initial content of the solids disc $M_{\rm p,ini}$, but the systems are now colour-coded by the four architecture classes. Here, we note a strong separation between the first two (only low-mass planets) and last two (with giant planets) classes of systems.

For the first two categories of systems, most of the planet masses is comprised of solids. Therefore, the masses in planets is generally a given fraction of the disc mass, between \SI{20}{\percent} and \SI{35}{\percent}, as mentioned. Usually, the whole inner disc up to \num{1} to \SI{2}{\au} is accreted by the protoplanets, some fraction beyond that up to \SI{\sim10}{\au}, while the outer part remain essentially intact, because the accretion timescales are too long. It should be taken into account that extending the (numerical) integration time (here \SI{100}{\mega\year}) in the simulations could further increase the efficiency for some systems, until ejection of planetesimals will start to limit further planet growth \citep{2004ApJIda1}. We also note that there is a trend between Class I and II. The lower-mass discs usually form Class I architectures, while more massive discs tend to result in Class II systems.

As discussed, discs with solids content $M_{\rm p,ini}$ above \num{150} to \SI{200}{\mearth} lead to the formation of giant planets. In this case, planets accrete significant envelopes, which strongly increases the conversion efficiency. However, we do not find a clear separation in the parameter space between Class III and IV architectures, which is interesting. We also searched the other initial conditions, including dust-to-gas ratio and photoevaporation rate, but did not find anything clearer than what is presented in Fig.~\ref{fig:sys_dsm_ptm}. The right panel shows that Class IV systems have larger disc masses than Class III systems, with median values of nearly \SI{440}{\mearth} and \SI{563}{\mearth}, respectively, although there is significant overlap. This leaves us to conclude that the distinction between the two classes is not only related to initial disc characteristics and that the chaotic nature of the dynamical interactions between the protoplanets also bears some responsibility. So for similar disc initial conditions, Class III or IV systems can arise, depending on the exact locations and timing of the moments the proto-giant planets in one system start runaway accretion (which depends for example on the exact initial positions of the embryos), and the associated presence or absence of strong gravitational interactions. This (quasi) chaotic nature is different than in a simpler model where only a single embryo per disc  is present, whose outcome is easy to predict \citep{2021AASchleckerB}. It should be noted that our full model used here is also deterministic, i.e., for given model initial conditions it always predicts exactly the same outcome, but small changes of the initial conditions may lead to very different outcomes because of N-body interactions.

Finally, examples of dynamically-active Class IV systems leading to the loss of most of the planets are visible in the bottom right corner of Fig.~\ref{fig:sys_dsm_ptm}'s left panel. Here close encounters lead to the loss of all giant planets and only lower-mass planets remain. This leads to the (rare) points with  high $M_{\rm p,ini}$ but (very) low $M_{\rm sys}$. If we were to include in the system mass also all planets that were ejected out into interstellar space or that collided with the host star (and not only the planets that actually exist at \SI{5}{\giga\year}), these points would join the other `normal' points at high $M_{\rm sys}$.

The classification by \citet{2023AAMishraA,2023AAMishraB} was applied to the same synthetic data as our approach. It is a complementary approach since, in contrast to our human-based approach (that stresses underlying physical processes and the emergence of planetary system architecture in time), it can be applied to observational data without knowledge of the formation history of the systems or the underlying formation processes. Visual comparison between the left panel of Fig.~\ref{fig:sys_dsm_ptm} and the bottom right panel of Fig.~2 of \citet{2023AAMishraB} reveals the nevertheless existing correspondence between our classification and theirs. We note that, broadly speaking, the `similar' class of \citet{2023AAMishraB} corresponds to our Class~I and~II. Also, our Class~III systems correspond to `mixed' systems and a part of the `ordered' ones, while Class~IV corresponds to the other part of `ordered' systems and the `anti-ordered' ones. The approximate match of the automated approach and the human classification for our multi-faceted data set give a promising outlook for a future holistic categorisation for observed systems. This should take into account planetary composition information as an additional formation tracer, both regarding the bulk and the atmospheric composition \citep{2022ApJMolliere}.

\subsection{Effect of the stellar environment}
\label{sec:res-env}

\begin{figure}
	\centering
	\includegraphics[width=0.495\textwidth]{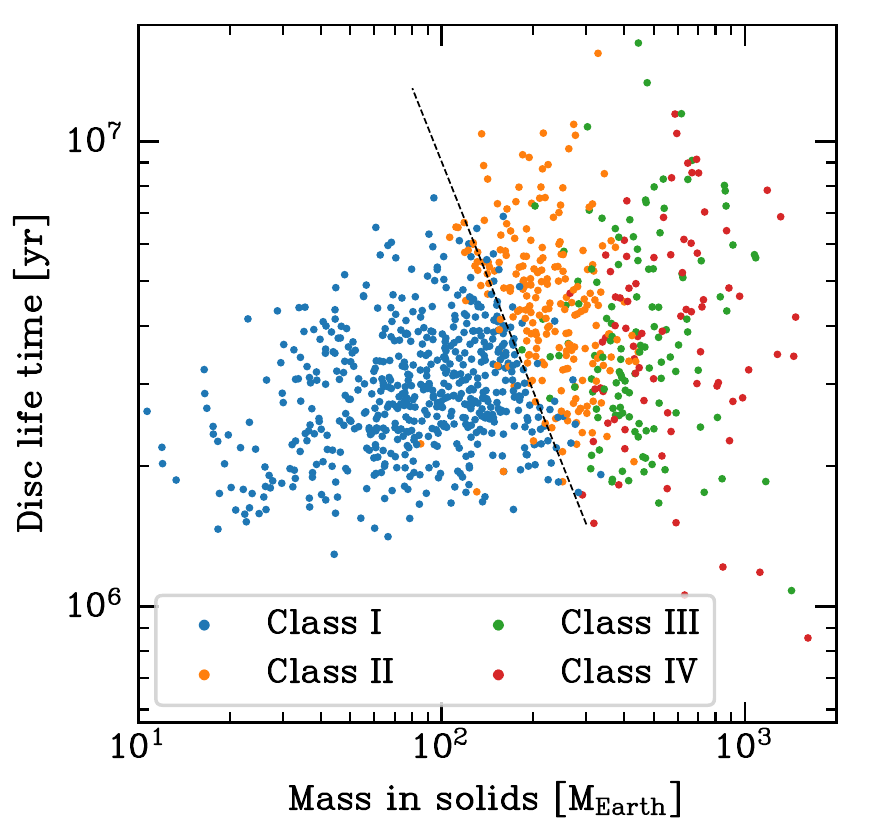}
	\includegraphics[width=0.495\textwidth]{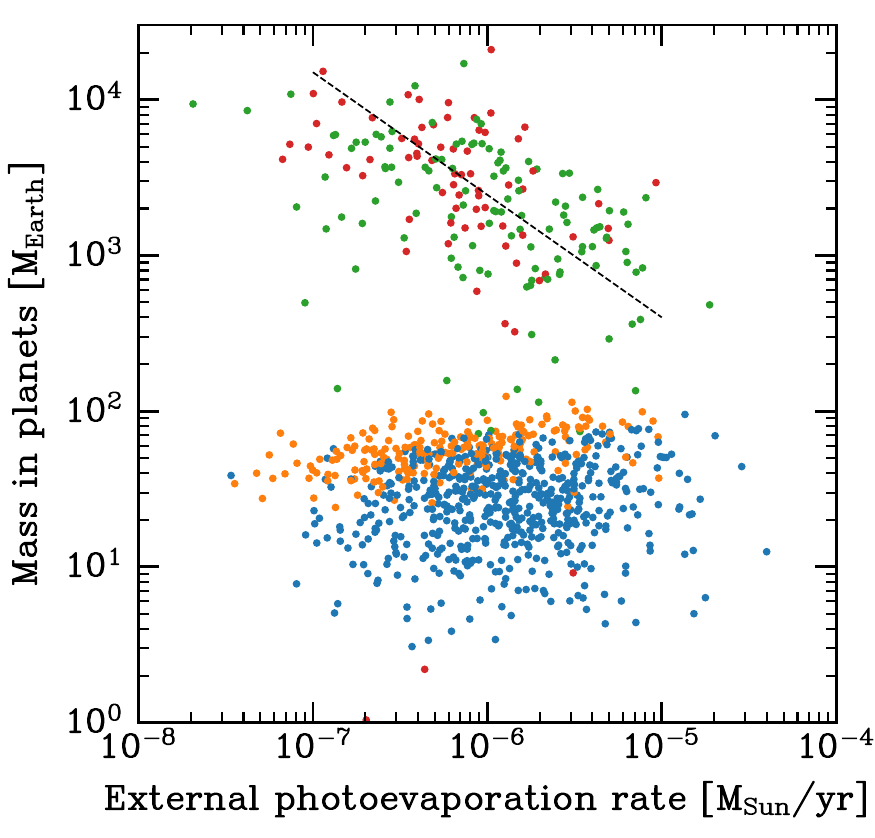}
	\caption{System architecture as function of the disc lifetime and external photoevaporation rate as proxies of the stellar environment. Left panel: architectural class as as a function of initial solids content of the disc $M_{\rm p,ini}$ and the gas disc's lifetime (which depends via external photoevaporation on the stellar environment). Right: total mass in planets per system  $M_{\rm sys}$ as function of the external photoevaporation rate. In the left panel, the dashed black line denotes roughly the separation between the conditions leading to Class~I and II systems. In the right panel, the dashed black line shows the anticorrelation between the mass in planets and the external photoevaporation rate in Class~III and~IV systems.}
	\label{fig:sys_env}
\end{figure}

The surrounding stellar environment drives the ambient field, which in turn drives the external photoevaporation of the protoplanetary disc. External photoevaporation will truncate the outer disc and reduce their lifetimes \citep{2018MNRASHaworth}. This in turn will affect planet formation, for instance by limiting the formation time or reducing the migration \citep{2022MNRASWinter}. Other effects of the environment include the heating of the disc \citep[e.g.]{2018MNRASNdugu} and stellar encounters \citep{2020MNRASStock}, but we do not include these effects in our model. Here, we focus on the effect of the external photoevaporation rate and discuss two cases where we find a correlation between the surrounding environment and either the class of system or the planet masses.

The left panel of Fig.~\ref{fig:sys_env} shows that the transition between Class~I and~II systems depends not only on the solids mass, but also the lifetime of the protoplanetary disc as illustrated by the dashed black line. The longer the disc lives, the less massive it needs to be for systems to be in Class II. There is the following reason for this: Class II systems occur when the planets get sufficiently massive to migrate substantially when the gas disc still has significant mass. Thus, the planets have to reach a mass of about \SI{10}{\mearth} (i.e. to reach the saturation and/or the equality mass) well before the dispersal of the gas disc. Now, the more massive the solid disc, the faster the planets grow. This means that short-lived discs needs to grow the massive icy planets fast, otherwise they will not have the time to migrate. In other words: at fixed $M_{\rm p,ini}$ in the range of about 100 to \SI{300}{\mearth}, short disc lifetimes lead to Class I systems, while long ones lead to Class II, because there needs to be sufficient time for orbital migration to occur to form a Class II system. A prime example of this effect can be seen by comparing the system in bottom right panel of Fig.~\ref{fig:class1examples} (Class~I) and the one on the top left panel of Fig.~\ref{fig:class2examples} (Class~II). Both systems have a similar initial solid disc mass (120 versus \SI{130}{\mearth}), but different gas disc masses and different disc lifetimes (\num{3.5} versus \SI{7.4}{\mega\year}). This results in terrestrial planets being able to survive in the former (although icy planets migrated to within \SI{1}{\au}) while they no longer exist in the latter. The link to the environment comes from the anticorrelation of disc lifetimes with external photoevaporation rate \citep{2022EPJPWinterHaworth}. Absent this, disc lifetimes are correlated to their initial mass. Thus, our model predicts that ordered and compositionally well-separated planets are more likely in dense clusters.

The second dependency on the environment we find is between the mass of $M_{\rm sys}$ for Class III and IV system, i.e., the mass of the giant planets and the external photoevaporation rate, as shown in the right panel of Fig.~\ref{fig:sys_env}. Here again, the reason is connected with the formation time of the planets. The external photoevaporation rate strongly affect the disc lifetimes, thus the time planets can accrete from the disc. The growth time of a massive core is related to the mass of the solids disc, thus relatively independent of the external photoevaporation rate. The cores will undergo runaway gas accretion at roughly the same times. However, external photoevaporation competes with planetary gas accretion. A similar correlation in observation was also obtained \citep{2020NatureWinter}.

\subsection{Total number of giant planets versus surviving and loss channels}

\begin{figure}
    \centering
    \includegraphics[width=0.495\textwidth]{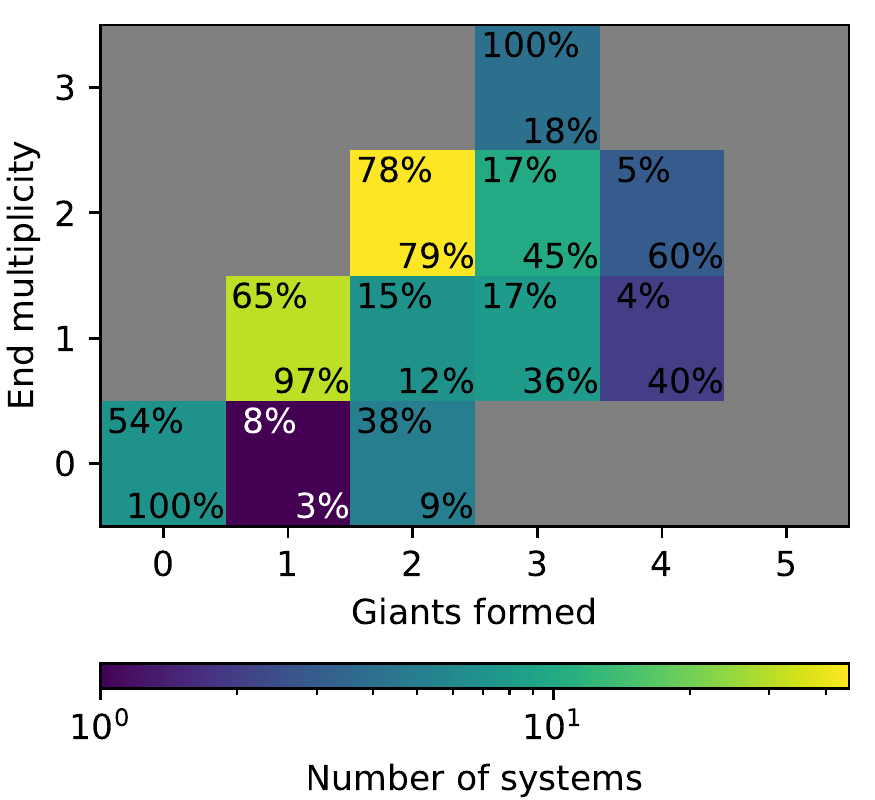}
    \includegraphics[width=0.495\textwidth]{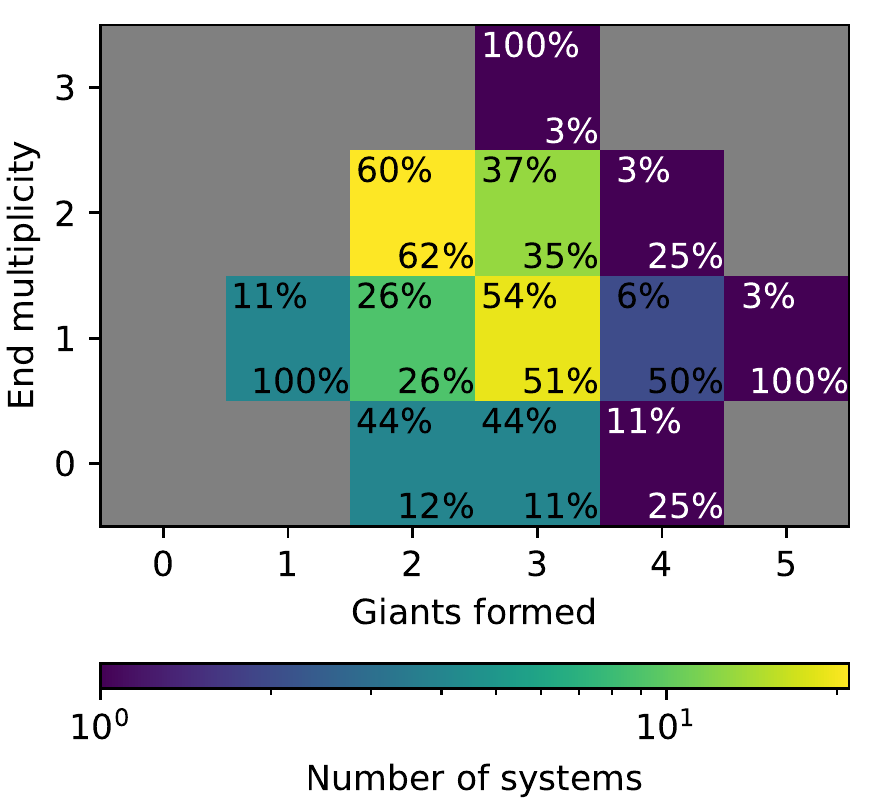}
    \caption{Correlation between the total and final number of giant planets in each systems for those of Class III (mixed systems; left) and Class IV (giants only; right). In each cell, the number of the top gives the percentage of systems with the same end multiplicity originate from systems that formed that number of giants (the sum of the top percentage on a row is unity, save for rounding errors) while the bottom number give the percentage of systems who formed the given number of giant planets ended with that multiplicity (the sum of the bottom percentage on a column is unity, save for rounding errors). The colour represent the underlying number of systems in each cell, on a logarithmic scale.}
    \label{fig:giant_multi_correl}
\end{figure}

As we saw, the formation of systems with (several) giant planets is often not a smooth process but may involve strong dynamical events: the final system architecture thus may not reflect the full picture of what happened during the formation stage. In addition, we did not find any clear separation in the initial conditions between systems in Classes~III and~IV. So, we want to check if the formation history can help us to  understand the final architecture. To this effect, we select one simple metric: the total number of giant planets that have been formed, no matter what they became (remained in the system till the end; accreted by another more massive giant; ejected into outer space; collided with the host star). We show the relations of the this total number versus the number of giants that remain till the end in Fig.~\ref{fig:giant_multi_correl} for two classes of systems: the mixed systems (Class III) and the ones with essentially only giants (Class IV). For completeness, we note that one system in Class II (migrated sub-Neptunes) has one giant planet at one point that was then accreted by the central star; otherwise none of these have or had giants.

The results show that most of the mixed systems (Class~III -- in total 123 systems) did not loose any giant planets, as one might expect. The two main situations are systems with two giants formed and two remaining (44 systems), then one giant formed and one remaining (31 systems). Accounting for the systems that formed and retained three or zero giants, there are in total 86 systems that did not loose any giant, compared to 37 that did loose at least one. The largest loss channel for these giants is giant impacts (about \SI{53}{\percent}) followed by mass loss from the envelope (meaning planets that are still present, but were classified as giant at one point during the formation and no longer are at the end, for example due to the ejection of the envelope after a giant impact; about \SI{33}{\percent}). Only about \SI{14}{\percent} of the giants that are lost are due to  ejections or collisions with the central star.

For the systems with only giant planets (Class~IV -- in total 80 systems), the story is quite different. Here, 23 systems did not loose any giant planets while 57 did loose at least one. Further, the distribution of giants lost across the different channels is different from Class III. About \SI{82}{\percent} of the losses are the result of ejections or collisions with the central star.

Thus, in total, nearly half of the systems that formed giant planets did loose at least one giant (94 out of 203). However, these are disproportionally found in systems that are left with giant planets only (Class~IV). On top of that, the loss mechanisms are different between the classes. The loss of giant planet by ejection or collision with the central star is usually sufficient to destabilise the inner system, so that the resulting planetary system ends up in Class~IV. Conversely, the loss of a giant planet by collision (a giant impact between giant planets) is in itself usually not sufficient to destabilise the system. Planetary systems who did not loose a giant planet or did so by collisions are more likely to be mixed systems (Class~III).

Nevertheless, the ejection of a giant planet is not a complete indicator of dynamical activity. We may give as example system \texttt{4} shown in Fig.~\ref{fig:class4examples}, which is classified as giants-only (Class~IV). It formed two giants yet did not loose either of them. Here, we have a pathway that the two giant planets interacted dynamically and that alone led to collisions and ejections in the inner part of the system.

\subsection{Effect of stellar metallicity}

\subsubsection{Planet eccentricities}

\begin{figure}
    \centering
    \includegraphics[width=0.495\textwidth]{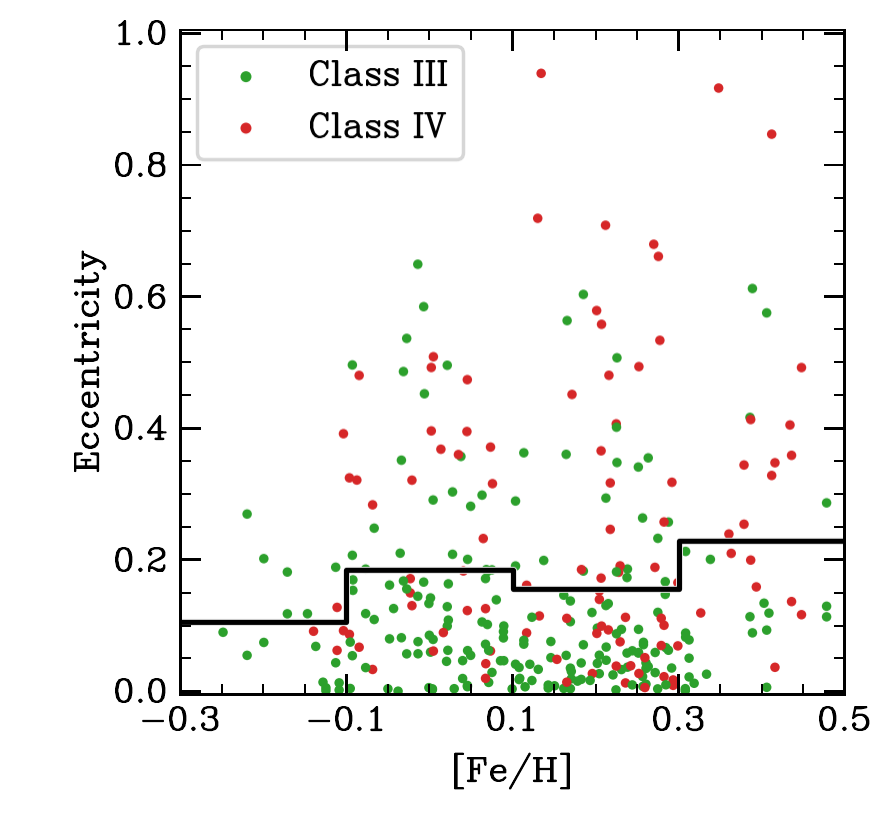}
    \includegraphics[width=0.495\textwidth]{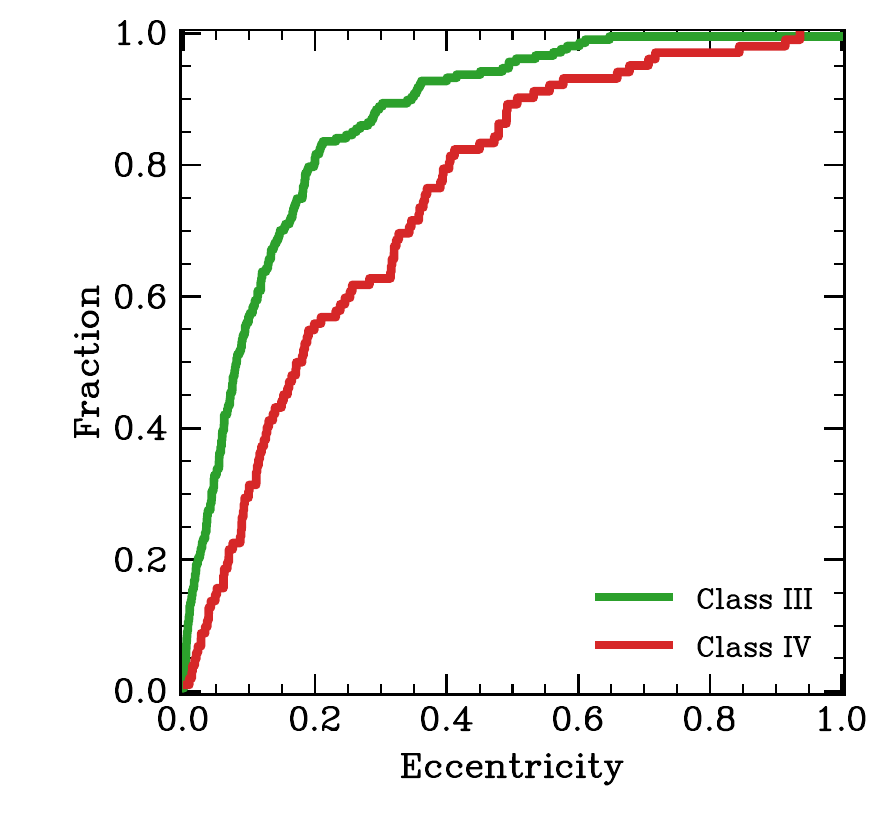}
    \caption{Eccentricity distribution of the giant planets (mass \SI{>100}{\mearth}). Left: eccentricity versus initial metallicity. The black line shows the mean value in each interval. Right: cumulative distribution of giant planets eccentricities in Class~III and Class~IV systems.}
    \label{fig:ecc_feh}
\end{figure}

There are several correlations between host star metallicity and exoplanet characteristics, as discussed in Sect.~\ref{sec:pps-exodem}. One is a positive trend between stellar metallicity and giant planet eccentricities \citep{2013ApJDawson,2018ApJBuchhave}. To check for this effect in our model, we provide in the left panels of Fig.~\ref{fig:ecc_feh} the eccentricities of giant planets as function of metallicity. We first see that overall giant planet eccentricities are rather low in our model, with a mean value of \num{0.17}. Still, we get a slight correlation between stellar metallicity and the mean eccentricity (represented by the black line).

In addition, we see a different trend in the eccentricities as function of the system class. The highly-eccentric planets ($e \gtrsim 0.6$) are mostly found in Class~IV systems and the mid-eccentricities ($0.6 \gtrsim e \gtrsim 0.3$) are also dominated by Class~IV systems, though a higher proportion of planets in Class~III systems is also present. Only in the low eccentricity regime ($0.3 \gtrsim e$) do planets in Class~III systems dominate. To support this, we provide in the right panel of Fig.~\ref{fig:ecc_feh} distinct cumulative functions for the giant planet eccentricities as function of class. There, we see that the eccentricity distribution is different between the two classes, with planets in Class~IV systems having overall larger eccentricities.

We also have a different distribution of stellar metallicities between Class~III and Class~IV systems. We already saw in Fig.~\ref{fig:sys_dsm_ptm} that Class~IV systems tended to have larger initial disc masses and this is also reflected in the metallicity distribution. The median metallicity of Class~III systems is \num{0.11} while for Class~IV systems it is \num{0.17}. Thus, our model can reproduce the general trend of increasing planet eccentricities with stellar metallicities, albeit eccentricities are overall low.

\subsubsection{Planet and system kinds}

\begin{figure}
    \centering
    \includegraphics{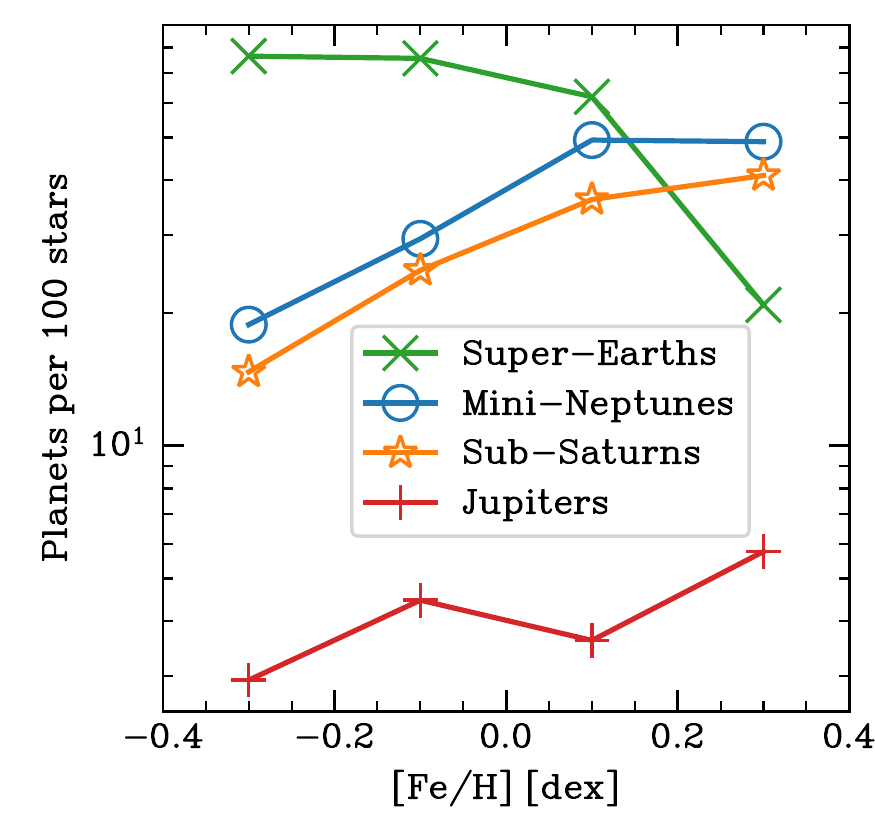}
    \includegraphics{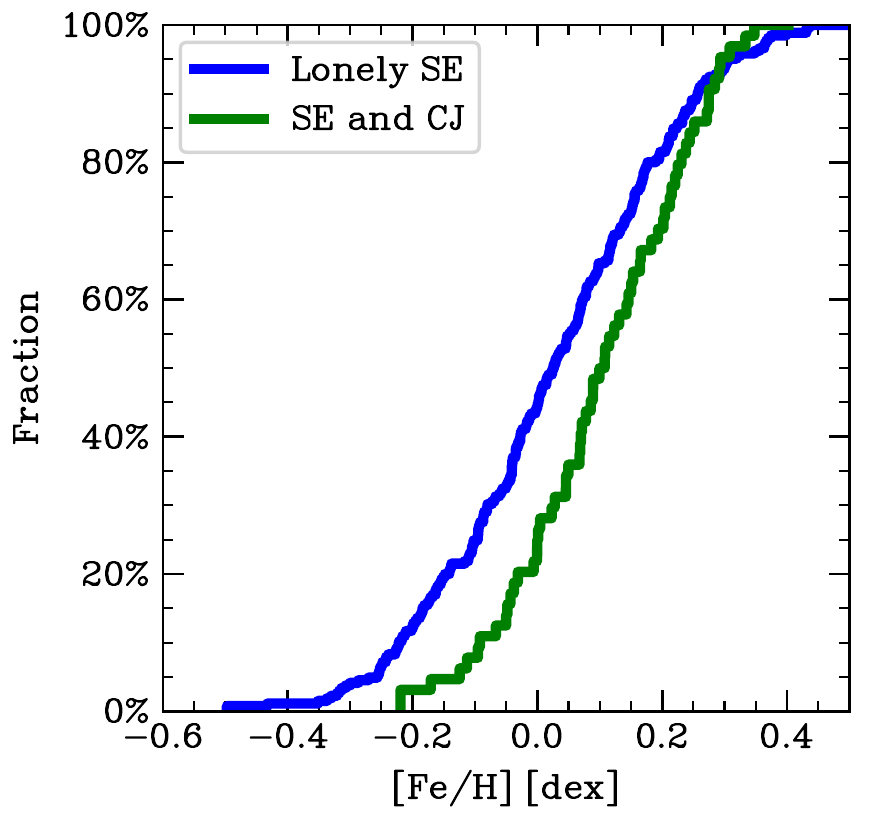}
    \caption{Metallicity dependence of certain planet and system kinds in the synthetic population, which can be compared to observations. Left: occurrence of planet classes at periods between \num{10} and \SI{100}{\day} (including an observability criterion similar to \citealp{2018AJPetigura}) as function of stellar metallicity (transit detections). Super-Earths are between \num{1.0} and \SI{1.7}{\rearth}, mini-Neptunes between \num{1.7} and \SI{4.0}{\rearth}, sub-Saturns between \num{4.0} and \SI{8.0}{\rearth}, and Jupiters above \SI{8.0}{\rearth}. Right: cumulative distribution of metallicities of systems harbouring an inner Super-Earth (SE; defined as distance between \num{0.02} and \SI{1}{\au} and a mass between \num{2} and \SI{30}{\mearth}) and with (without) outer giant planet -- or cold Jupiter (CJ) -- in green (blue), defined as distance between \num{0.23} and \SI{10}{\au} and mass between \num{30} and \SI{6000}{\mearth}. Only planets with a radial velocity semi-amplitude $K>\SI{2}{\meter\per\second}$ are accounted for.}
    \label{fig:types_metal}
\end{figure}

There are observational correlations between kinds of either planets or system architectures and stellar metallicity, both in transit and radial velocity surveys. To assess our model against such trends, we perform two comparisons. First, we compare the metallicity effect of different kind of planets following \citet{2018AJPetigura}. Here we select planets whose orbital period is between \num{10} and \SI{100}{\day}, apply an selection similar to transit surveys, as described in \citet{2021AAEmsenhuberB}, and categories the planets according to their radii following the definitions of \citealp{2018AJPetigura}: super Earths from \num{1.0} to \SI{1.7}{\rearth}, mini-Neptunes from \num{1.7} to \SI{4.0}{\rearth}, sub-Saturns from \num{4.0} to \SI{8.0}{\rearth}, and Jupiters above \SI{8.0}{\rearth}. The occurrence of small-radii planets, `super Earths', show an anti-correlation with stellar metallicity. For comparison, \citet{2018AJPetigura} found that the occurrence of the same planets (panel b of their Fig.~10) has overall no significant correlation with metallicity. However, the occurrence rate of the largest metallicity bin (0.2--0.4) is lower than the others, as in our simulations. Our finding is also consistent with the occurrence of low-mass planets reported by \citet{2021AAEmsenhuberB}, where systems having Earth-like (mass from \num{0.5} to \SI{2}{\mearth}) planets within \SI{1}{\au} have a median metallicity of \num{-0.09} (the same value for the population presented here is \num{-0.11}). All other categories, from mini-Neptunes to Jupiters show a positive correlation between occurrence rate and metallicity. For the mini-Neptunes and sub-Saturns this is consistent with the trend found by \citet{2018AJPetigura}, while the authors have been unable to find any trend for Jupiters from the Kepler mission, as there were only few such planets observed. Our model is thus able to reproduce the general trends correlation with metallicity found in the Kepler survey.

87\% of super Earths are found in Class~I systems, whose occurrence rate is anticorrelated with metallicities (Fig.~\ref{fig:sys_class_hist}, centre panel). This, combined with their total absence from Class~IV systems, explains the trend of anti-correlation between the occurrence of super Earths with stellar metallicity. Mini-Neptune planets originate principally from Class~II (49\%) and Class~I (40\%) systems. As the occurrence of Class~II systems is positively correlated with stellar metallicity, this explains the shift in the metallicity correlation between super Earths and mini-Neptunes. From this discussion, we also see that Class~I systems may produce large planets (both in mass and radii) which enter the mini-Neptune category. Having them originate more from Class~II systems would be a way to reinforce the metallicity correlation.

A second comparison can be done on the correlation between inner super Earths and distant giant planets (or cold Jupiters). Such a correlation was observed \citep{2018AJZhuWu,2019AJBryan} and the metallicity effect of these systems was studied by \citet{2022ApJSRosenthal}. \citet{2021AASchleckerA} determined there was a correlation between the occurrence of super Earths and cold Jupiters in the same model as presented in this work (although assuming \SI{20}{\mega\year} for the formation stage), though weaker than in observations. Here we extend the comparison for the trend in metallicity between systems hosting only inner super Earths and both super Earths and cold Jupiters in radial velocity survey. For this, we first restrict the synthetic planets to those with a radial velocity semi-amplitude $K>\SI{2}{\meter\per\second}$, as in \citet{2021AASchleckerA}. We use the same definitions as \citet{2022ApJSRosenthal}, name super Earths have distances between \num{0.02} and \SI{1}{\au} and masses between \num{2} and \SI{30}{\mearth} while cold Jupiters have distances between \num{0.23} and \SI{10}{\au} and masses between \num{30} and \SI{6000}{\mearth}. The cumulative distributions of metallicity for system hosting only super Earths and both super Earths and cold Jupiters are shown with the blue and green curves in the right panel of Fig.~\ref{fig:types_metal}, respectively. The \textit{p-value} of a Kolmogorov-Smirnov (KS) test between the two distributions to assess whether they are drawn from the same sample yields \num{3.7e-3}, implying that the two distributions are distinct. Our results are compatible with those \citet{2022ApJSRosenthal} in that 1) systems with only super Earths have lower metallicities than those with both super Earths and cold Jupiters, 2) a statistical test assesses that the two groups have a distinct underlying distribution, and 3) the metallicity distributions are broadly similar. In \citet{2022ApJSRosenthal}, the metallicity of systems with only super Earths spans roughly \num{-0.4} to \num{0.3} with most of the systems between \num{-0.2} and \num{0.2}, while the metallicity of systems with both super Earths and cold Jupiters spans roughly \num{0.05} to \num{0.35} (their Fig.~8). In our model, the values are similar although the the metallicities of the latter type are shifted towards lower values, with values mainly between \num{-0.1} and \num{0.3}. We find that \SI{77}{\percent} of the systems with both super Earths and cold Jupiters have $[\mathrm{Fe/H}]>0$, and \SI{41}{\percent} of them have $[\mathrm{Fe/H}]>0.15$. Thus, while the systems containing both kinds of planets are not as strongly biased towards large metallicities in our synthetic population compared to what \citet{2022ApJSRosenthal} obtained, we nonetheless obtain a bias.

When comparing these types of systems with our classes, we find that systems with only detectable super Earths are \SI{77}{\percent} from Class~II, \SI{20}{\percent} from Class~I and \SI{3}{\percent} from Class~III. Systems with both super Earths and cold Jupiters are \SI{94}{\percent} from Class~III, with contributions from all others categories. We note that even a system of Class~I is included here; this is because that system contains a rocky planet of \SI{18}{\mearth} at \SI{0.15}{\au} and an icy planet of \SI{36}{\mearth} at \SI{0.39}{\au}. Although the planet are massive for their kind (and hence enter the definitions of super Earth and cold Jupiters of \citealp{2022ApJSRosenthal}), the formation pattern of the system is similar to that of Class~I with only a small effect of migration and was thus categorised as such. Thus, the difference in metallicities between super Earth systems and with or without cold Jupiters is consistent with Class~III systems having overall larger metallicities than Class~II systems.

We find different metallicity effects for planets defined as `super Earths' in transit or radial velocity surveys. The super Earths according to the definition of \citet{2018AJPetigura} show an anticorrelation, or a negative metallicity effect. Conversely, using the definitions of the radial velocity survey of \citet{2022ApJSRosenthal}, the median metallicity of systems hosting super Earths is \num{0.05}, thus showing a positive metallicity effect. The difference is due to \SI{75}{\percent} of the defined as super Earths by \citet{2022ApJSRosenthal} having radii between \num{1.7} and \SI{4.0}{\rearth}, the range defined as `mini-Neptunes' by \citet{2018AJPetigura}. We thus caution that there is an inconsistency between the naming of `super Earths' in transit and radial velocity, and that `super Earths' in radial velocity are similar to `mini-Neptunes' in transit surveys, which both show a positive metallicity effect.

We further find that the median metallicity of systems where a giant planet has been accreted by the central star is \num{0.15}. This is an indication that our model could reproduce the metallicity effect of the high-eccentricity migration channel of hot-Jupiters \citep{2013ApJDawson,2018ARAADawson}. However, we cannot fully confirm this because the model does not account for tidal circularisation.

\section{Discussion and conclusion}

In this work, we discussed how planetary population synthesis can be used to infer how planetary formation may work although we have at the moment only few direct observations of forming planets. The general idea is to find a theoretical end-to-end model that can reproduce the observed planet population from the diversity of known protoplanetary discs in a quantitative statistical way. The underlying theoretical model has to be global, so that it includes the essence of all (currently known) important governing physical processes and treat their interactions. We can also use such a global model to investigate the physical processes that lead to different classes of architectures of synthetic planetary systems and how the disc and stellar environmental properties affects the outcome. Hence, we classified the \num{1000} synthetic systems from the nominal NGPPS population of \citet{2021AAEmsenhuberB} based on classes of formation pathways seen in the simulations and resulting planetary system architectures. This approach stresses the temporal emergence of system architectures and the underlying physical processes included in the theoretical model. This is in contrast to the approach of \citet{2023AAMishraA,2023AAMishraB}, who classified the systems based on the observable present day properties of the final systems.

We find four classes of formation and architecture: (compositionally) ordered Earths and ice worlds systems (Class~I), migrated sub-Neptunes systems (Class~II), mixed systems with inner low-mass and giant planets (Class~III), and systems of dynamically-active giants (Class~IV). Then, three different types of analysis of these classes are presented: first, using (analytical) mass scales, namely the growth, equality, saturation, Goldreich, and the critical core mass which can be used to understand which physical process is mainly responsible for shaping the final systems architecture and the characteristic mass scale in a system. Second we studied the disc initial conditions and the stellar environment that lead to the formation of systems in the four classes. Third, we assess how the model reproduces the observational trends with stellar metallicity on planet eccentricities and the  occurrence of certain kinds of planet and architectures.

The compositionally ordered Earths and ice worlds systems, which consist of low-mass rocky planets inside, and icy ones outside, form from the lowest-mass discs mostly by in-situ accretion of planetesimals followed by giants impacts among the protoplanets. The relevant mass scaling for the final planet masses is the Goldreich mass, which applies to systems ending with a giant impact phase. Intermediate-mass discs lead to migrated sub-Neptune systems, where sub-Neptunian planets migrate all the way to the inner disc edge. The relevant mass scaling is here the equality mass (where the accretion and migration timescales are the same) and/or the saturation mass (where planets are released from Type I migration traps because the positive corrotation torque saturates). It is interesting to note that the disc solid mass required to transition from Class~I to Class~II depends on the disc lifetime. This is because more massive discs form sub-Neptunes more rapidly and then undergo gas-driven migration towards the inner region.

When assessing the correlation between planet kind and metallicity, we find that the observed trends are reproduced, although overall super Earths and sub-Neptunes are found in lower-metallicity environment than in observations. In our synthetic populations, a significant number of observable sub-Neptunes originate from Class~I systems. This tends to favour systems with lower metallicities. Having these planets originate predominantly from Class~II systems would favour higher metallicities, providing a better correspondence between the model and observations. Whether this is possible should be investigated in the future.

The other two classes correspond to systems with giants planets that have (Class~III) or do not have (Class~IV) inner lower-mass companions. Here, the relevant formation pathway is that the equality mass must be of the order of \SI{10}{\mearth}, which corresponds to the critical core mass, to allow for strong gas accretion while the planets migrate. The distinction between the two classes is whether some strong dynamical instability has taken place between the giants, which lead to the loss of the inner companions. While these two categories together are well separated from the others in the space of initial conditions (Fig.~\ref{fig:sys_dsm_ptm}), there is no such clear separation in the initial disc properties (like initial mass of solids) between the systems in Class~III versus Class~IV. Thus, we attribute the discerning factor mainly to the stochasticity of the \textit{N}-body interactions: for some initial placings of the embryos, the forming giants do undergo dynamical instabilities, for others not. There are nevertheless some trends, with Class~IV systems forming preferentially from higher-mass discs (Fig.~\ref{fig:sys_dsm_ptm}) and where photoevaporation is lower and thus disc lifetimes are longer (Fig.~\ref{fig:sys_env}).

We further investigate whether the total number of giant planets that ever existed in a system can explain the distinction between Classes~III and~IV. We find that the loss of a giant planet due to an ejection or a collision with the central star (as the result of a close encounters between two giant planets) will likely result in the destabilisation of the inner system, thereby resulting in a system with only giant planets. However, if the two giant planets collide rather than scatter themselves, this is not sufficient to destabilise the inner system. Thus, the stochastic nature of close encounters plays an important role in the distinction between Class~III and~IV, rather than only the initial conditions.

The discovery space of extrasolar planet detection methods expands continuously. Several upcoming missions and instruments like GAIA, NIRPS, PLATO, Roman, or Ariel will foster this trend even more, and allow to build observationally a more complete statistical picture of the demographics of extrasolar planets. The times are therefore promising for a method like planetary population synthesis which can take full advantage of the wealth of all these new observational constraints. It will, however, be important to compare the same model with all different constraints and methods simultaneously to uncover the shortcomings in our theoretical understanding, and to avoid building models to reproduce only a certain observation. Furthermore, it is time to conduct comparisons of models and observations in a quantitative, and no longer a qualitative fashion only, given that the observational data is fully on a level to allow for this. Combined with direct observations of ongoing planet formation, this should make it possible to markedly improve our understanding of the planet formation process in the coming decade.

\section*{Acknowledgements}

We thank the anonymous reviewer for the helpful comments and suggestions.
This work was funded in part by the Deutsche Forschungsgemeinschaft (DFG, German Research Foundation) - 362051796. It has been carried out partially within the framework of the NCCR PlanetS supported by the Swiss National Science Foundation under grants 51NF40\_182901 and 51NF40\_205606.
C.M. acknowledges the funding from the Swiss National Science Foundation under grant 200021\_204847 `PlanetsInTime'.
Most plots shown in this work were generated using matplotlib \citep{2007CSEHunter}.

\section*{Data availability}

The datasets analysed during the current study are available from the corresponding author on reasonable request.

\bibliographystyle{apalike}
\bibliography{manu}

\end{document}